\documentclass{vldb}

\usepackage{booktabs} 

\usepackage{graphicx}

\usepackage{epstopdf}
\usepackage[font={small}]{caption}
\usepackage{subcaption}
\usepackage{fixltx2e}
\usepackage{balance}
\usepackage[linesnumbered,ruled,vlined,noend]{algorithm2e}
\usepackage{textcomp}
\usepackage{amsmath}

\usepackage{amssymb}  
\usepackage{amsfonts}
\usepackage{color}

\usepackage{xspace}
\usepackage{multirow}

\usepackage{stfloats}

\usepackage{footnote}
\makesavenoteenv{tabular}
\makesavenoteenv{table}

\usepackage{slashbox}

\usepackage{enumitem}

\usepackage{amssymb}

\newtheorem{exmp}{\textbf{Example}}
\newtheorem{theorem}{\textbf{Theorem}}
\newtheorem{proposition}{\textbf{Proposition}}
\newtheorem{defn}{\textbf{Definition}}

\newcommand{\ignore}[1]{}

\newcommand{\stitle}[1]{\vspace{1ex} \noindent{\bf #1}}

\DeclareMathAlphabet{\mathpzc}{OT1}{pzc}{m}{it}

\newcommand{\Pads}{{{\tt RADS}}\xspace}
\newcommand{\Mur}{{\textbf{R-Meef}}\xspace}

\newcommand{\BD}{{{BD}}\xspace}
\newcommand{\QD}{{Span}\xspace}
\newcommand{\sme}{{SM-E}\xspace}

\newcommand{\decomp}{{\mathcal{DE}}\xspace}
\newcommand{\joinplan}{{{PL}}\xspace}
\newcommand{\regiongroup}{{{RG}}\xspace}

\newcommand{\trienode}{{\mathcal{N}}\xspace}

\newcommand{\sPattern}{P \xspace}

\newcommand{\invTrie}{{\mathcal{ET}}\xspace}

\newcommand{\eTrie}{{\mathcal{ET}}\xspace}

\newcommand{\mds}{{{MCDS}}\xspace}

\newcommand{\evi}{EVI\xspace}

\newcommand{\pNode}{parentNode\xspace}

\newcommand{\tNode}{leaf node\xspace}

\newcommand{\ec}{{{EC}}\xspace}
\newcommand{\ecsb}{{{$\widetilde{\mathbb{R}}_{G_t}(P)$}}\xspace}

\newcommand{\ecsp}[1]{{{{R}_{G_t}({#1})}}\xspace}

\newcommand{\piv}{{piv}\xspace}
\newcommand{\lf}{{LF}\xspace}

\newcommand{\ei}{{I}\xspace}
\newcommand{\cds}{{{CDS}}\xspace}
\newcommand{\mcds}{{{MCDS}}\xspace}
\newcommand{\mlst}{{{MLST}}\xspace}
\newcommand{\psgl}{{{\tt PSgL}}\xspace}
\newcommand{\twintwig}{{{\tt TwinTwig}}\xspace}
\newcommand{\seed}{{{\tt SEED}}\xspace}
\newcommand{\et}{{embedding trie}\xspace}

\newcommand{\qiao}{{\tt Crystal}\xspace}

\newcommand{\rmeef}{{\textit{\textbf{r}egion-grouped \textbf{m}ulti-round \textbf{e}xpand v\textbf{e}rify $\&$ \textbf{f}ilter}}\xspace}

\date{}

\begin{document}

\title{Fast and Robust Distributed Subgraph Enumeration}
\numberofauthors{4} 
%

\author{
\alignauthor
Xuguang Ren \\
       \affaddr{Griffith University, Australia}\\
       \email{x.ren@griffith.edu.au}
\alignauthor
Junhu Wang \\
       \affaddr{Griffith University, Australia}\\
       \email{j.wang@griffith.edu.au}
\and
\alignauthor
Wook-Shin Han \\
       \affaddr{POSTECH, Public of Korea}\\
       \email{wshan@dblab.postech.ac.kr}
\alignauthor
Jeffrey Xu Yu \\
       \affaddr{The Chinese University of Hong Kong}\\
       \email{yu@se.cuhk.edu.hk}
}

\maketitle 
\thispagestyle{empty}
\pagestyle{empty} 

\begin{abstract}
We study the classic subgraph enumeration problem under distributed settings. Existing solutions either suffer from severe memory crisis or rely on large indexes, which makes them impractical for very large graphs. Most of them follow a synchronous model where the performance is often bottlenecked by the machine with the worst performance. Motivated by this, in this paper, we propose \Pads, a \textbf{R}obust  \textbf{A}synchronous  \textbf{D}istributed  \textbf{S}ubgraph enumeration system. \Pads first identifies results that can be found using single-machine algorithms. This strategy not only improves the overall performance but also reduces network communication and memory cost.
Moreover, \Pads employs a novel \rmeef framework which does not need to shuffle and exchange the intermediate results, nor does it need to replicate a large part of the data graph in each machine. This feature not only reduces network communication cost and memory usage, but also allows us to adopt simple strategies for memory control and load balancing, making it more robust.
Several heuristics are also used in \Pads to further improve the performance.
Our experiments verified the superiority of \Pads to state-of-the-art subgraph enumeration approaches.

\end{abstract}


\section{Introduction}
\label{sec:introduction}

Subgraph enumeration\ignore{, also known as subgraph listing,} is the problem of finding all occurrences of a query graph in a data graph.
Its solution is a basis for many other algorithms and it finds numerous applications. This problem has been well studied under single machine settings \cite{turboIso}\cite{boostIso}. However in the real world, the data graphs are often fragmented and distributed across different sites. This phenomenon
highlights the importance of distributed systems of subgraph enumeration. Also, the increasing size of modern graph makes it hard to load the whole graph into memory, which further strengthens the requirement of distributed subgraph enumeration.

In recent years, several approaches and systems have been proposed \cite{DBLP:conf/icde/AfratiFU13,DBLP:conf/sigmod/ShaoCCMYX14,DBLP:journals/pvldb/LaiQLC15,DBLP:journals/pvldb/LaiQLZC16,DBLP:conf/sigmod/FanXWYJZZCT17,DBLP:conf/sigmod/FanLLXYYX18}.
However,  existing systems either need to exchange large intermediate results (e.g., \cite{DBLP:journals/pvldb/LaiQLC15},\cite{DBLP:journals/pvldb/LaiQLZC16} and \cite{DBLP:conf/sigmod/ShaoCCMYX14}), or  copy and replicate large parts of the data graph on each machine (e.g., \cite{DBLP:conf/icde/AfratiFU13} and \cite{DBLP:conf/sigmod/FanXWYJZZCT17,DBLP:conf/sigmod/FanLLXYYX18}), or rely on heavy indexes (e.g., \cite{DBLP:journals/pvldb/QiaoZC17}). Both exchanging and caching large intermediate results and exchanging and caching large parts of the data graph will cause heavy burden on the network and on memory, in fact, when the graphs are large these systems tend to crash due to memory depletion.
In addition, most of the current systems are synchronous, hence they suffer from {\em synchronization delay}, that is, the machines must wait for each other for the completion of certain processing tasks, making the overall performance equivalent to that of the slowest machine.
More details about existing work can be found in Section~\ref{sec:relatedWork}.

%

It is observed in previous work \cite{DBLP:journals/pvldb/LaiQLZC16,DBLP:journals/pvldb/QiaoZC17} that when the data graph is large, the number of intermediate results can be huge, making the network communication cost a bottleneck and causing memory crash. On the other hand, systems that rely on replication of large parts of the data graph or heavy indexes are impractical for large data graphs and low-end computer clusters.  In this paper,  we present \Pads, a \textbf{R}obust \textbf{A}synchronous \textbf{D}istributed \textbf{S}ubgraph enumeration system. Different from previous work, our system does not need to exchange intermediate results or replicate large parts of the data graph.  It does not rely on heavy indexes or suffer from synchronization delay. Our system is also more robust due to our memory control strategies and easy for load balancing.  

To be specific, we make the following contributions:

\renewcommand\labelenumi{(\arabic{enumi})}
\renewcommand\theenumi\labelenumi
\begin{enumerate}
\item We propose a novel distributed subgraph enumeration framework, where the machines do not need to exchange intermediate results, nor do they need to replicate large parts of the data graph.  
 
\item We propose a method to identify embeddings that can be found on each local machine independent of other machines, and use single-machine algorithm to find them.  This strategy not only improves the overall performance, but also reduces network communication and memory cost.
    
\item We propose effective memory control strategies to minimize the chance of memory crash, making our system more robust. Our strategy also facilitates workload balancing.

\item We propose optimization strategies to further improve the performance. These include (i) a set of rules to compute an efficient execution plan, (ii) a dynamic data structure to compactly store intermediate results.   


%

\item  We conduct extensive experiments which demonstrate that our system is not only significantly faster than existing solutions \footnote{Except for some queries using \cite{DBLP:journals/pvldb/QiaoZC17}, which relies on heavy indexes.}, but also more robust.

\end{enumerate}

%
%
%
%

\stitle{Paper Organization}
In Section~\ref{sec:Preliminary}, we present the preliminaries. In Section~\ref{sec:outline}, we present the architecture and framework of our system \Pads. In Section~\ref{sec:executionPlan}, we present algorithms for computing the execution plan. In Section~\ref{sec:embeddingTrie}, we present the embedding trie data structure to compress our intermediate results. Our memory control strategy is given in Section~\ref{sec:memoryControl}. We present our experiments in Section~\ref{sec:experiment}, discuss related work in Section~\ref{sec:relatedWork} and conclude the paper in Section~\ref{sec:conclusion}. Some proofs, detailed algorithms and auxiliary experimental results are given in the appendix.

\section{Preliminaries}
\label{sec:Preliminary}

{\stitle{Data Graph \& Query Graph}}  Both the data graph and query graph (a.k.a query pattern) are assumed to be unlabeled, undirected, and connected graphs.  We use $G$ = ($V_G$, $E_G$) and $P$ = ($V_P$, $E_P$) to denote the data graph and query graph respectively, where $V_G$ and $V_P$ are the vertex sets,  and $E_G$ and $E_P$ are the edge sets.  We will use {\em data} (resp. {\em query}) {\em vertex} to refer to vertices in the data (resp. query) graph.
Generally, for any graph $g$, we use $V_g$ and $E_g$ to denote its vertex set and edge set respectively, and for any vertex $v$ in $g$, we use $adj(v)$ to denote $v$'s neighbour set in $g$ and use $deg(v)$ to denote the degree of $v$.

{\stitle{Subgraph Isomorphism}} Given a data graph $G$ and a query pattern $P$, $P$ is subgraph isomorphic to $G$ if  there exists an injective function  $f$: $V_P$   $\rightarrow$  $V_G$  such that for any edge ($u_{1}$, $u_{2}$) $\in$ $E_{P}$,  there exists an edge ($f$($u_{1}$), $f$($u_{2}$)) $\in$ $E_{G}$. The injective function is also known as an  \emph{embedding} of $P$ in $G$ (or, from $P$ to $G$), and it can be represented as a set of vertex pairs ($u$, $v$) where $u$ $\in$ $V_{P}$  is mapped to $v$ $\in$ $V_{G}$. We will use $\mathbb{R}_{G}(P)$ to denote the set of all embeddings of  $P$ in  $G$.

The problem of subgraph enumeration is to find the set $\mathbb{R}_{G}(P)$. In the literature, subgraph enumeration is also referred to as subgraph isomorphism search  \cite{DBLP:journals/pvldb/LeeHKL12}\cite{turboIso}\cite{boostIso} and subgraph listing \cite{DBLP:conf/sigmod/KimLBHLKJ16}\cite{DBLP:conf/sigmod/ShaoCCMYX14}.

{\stitle{Partial Embedding}}
A \emph{partial embedding} of graph $P$ in graph $G$ is an embedding in $G$ of a vertex-induced subgraph of $P$. A partial embedding is a full embedding if the vertex-induced subgraph is $P$ itself.

{\stitle{Symmetry Breaking}}
A symmetry breaking technique based on automorphism is conventionally used to reduce duplicate embeddings \cite{grochow2007network}. As a result the data vertices in the final embeddings should follow a preserved order of the query vertices. We apply this technique in this paper by default and we will specify the preserved order when necessary.

%

{\stitle{Graph Partition \& Storage}}
Given a data graph $G$ and $m$ machines $\{M_1, \dots, M_m\}$ in a distributed environment, a partition of $G$ is denoted $\{G_1, G_2, \dots, G_m \}$  where $G_t$ is the partition located in the $t^{th}$ machine $M_{t}$.
In this paper, we assume each partition is stored as an adjacency-list. For any data vertex $v$, we assume its adjacency-list is stored in a single machine $M_t$ and we say $v$ is owned by $M_t$ (or resides in $M_t$).
We call $v$ a foreign vertex of $M_t$ if $v$ is not owned by $M_t$.
We say a data edge $e$ is owned by (or resides in) $M_t$ (denoted as $e\in E_{G_t}$) if either end vertex of $e$ resides in $M_{t}$.
Note that an edge can reside in two different machines.

For any $v$ owned by $M_t$, we call $v$ a {\em border} vertex if any of
its neighbors is owned by other machines  than $M_t$.
Otherwise we call it  a {\em non-border} vertex. We use $V^b_{G_t}$ to denote the set of all border vertices in $M_t$.

\section{RADS Architecture}
\label{sec:outline}

In this section, we first present an overview of the architecture of $\Pads$, followed by the $\Mur$ framework of $\Pads$. We give a detailed implementation of $\Mur$ in Appendix~\ref{sec:rmeefImplementaion}.

\subsection{Architecture Overview}
\label{subsec:architecture}

\begin{figure}[htbp!]
\small
\centering
\includegraphics[width=0.45\textwidth]{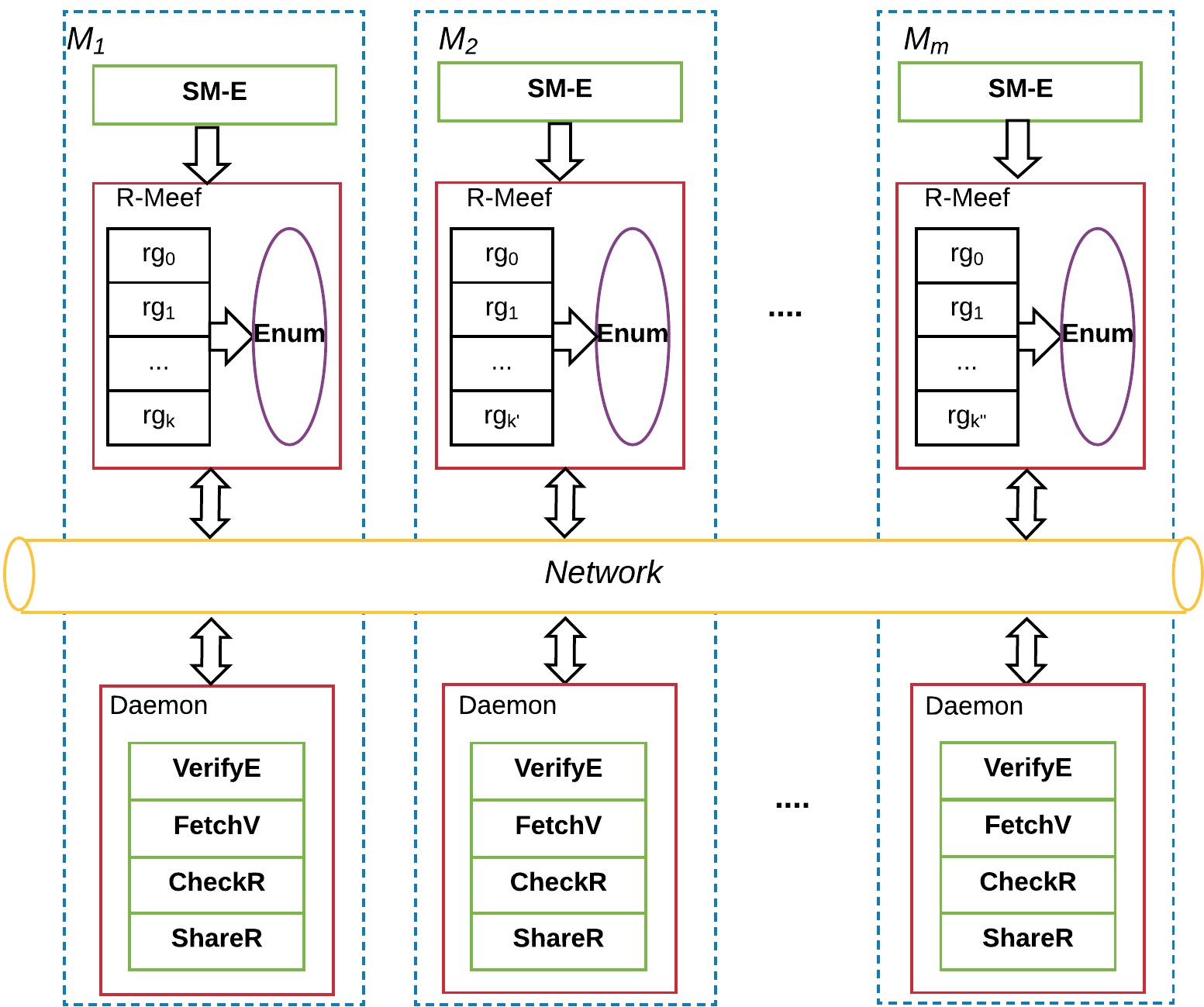}
\caption{RADS Architecture}
\label{fig:architecture}
\end{figure}

The architecture of $\Pads$ is shown in Figure~\ref{fig:architecture}.
{Given a query pattern $P$, within each machine, $\Pads$ first launches a process of single-machine enumeration (\sme) and a daemon thread, simultaneously. After \sme finishes, $\Pads$ launches a $\Mur$ thread subsequently. Note that the $\Mur$ threads of different machines may start at different time.}

\begin{itemize}[leftmargin=*]

\item \textbf{Single-Machine Enumeration}
The idea of \sme is to try to find a set of local embeddings using a single-machine algorithm, such as TurboIso\cite{turboIso}, which does not involve any distributed processing. The subsequent distributed process only has to find the remaining embeddings.  
This strategy can not only boost the overall enumeration efficiency but also significantly reduce the memory cost and communication cost of the subsequent distributed process.
{Moreover the local embeddings can be used to estimate the space cost of a {\em region group}, which will help to effectively control the memory usage (to be discussed in Section~\ref{sec:memoryControl}).}

We first define the concepts of {\em border distance} and {\em span}, which will be used to identify embeddings that can be found by \sme.

\begin{defn}[Border Distance]
\label{defn:borderDia}
Given a graph partition $G_t$ and data vertex $v$ in $G_t$, the {\em border distance} of $v$ w.r.t $G_t$,  denoted as $\BD_{G_t}(v)$, is the minimum shortest distance between $v$ and any border vertex of $G_t$, that is
\begin{equation}
\BD_{G_t}(v) = \min \limits_{v' \in V_{G_t}^b} dist(v, v')
\end{equation}
where $dist(v, v')$ is the shortest distance between $v$ and $v'$.
\end{defn}

\begin{defn}[Span]
\label{defn:queryDia}
Given a query pattern $P$, the {\em span} of query vertex $u$, denoted as $\QD_{P}(u)$, is the maximum shortest distance between $u$ and any other vertex of $P$, that is
\begin{equation}
\QD_{P}(u) = \max \limits_{u' \in V_P} dist(u, u')
\end{equation}
\end{defn}


\begin{proposition}
\label{lem:diaEmbed}
Given a data vertex $v$ of $G_t$ and a query vertex  $u$ of $P$, if $ \QD_{P}(u)\leq \BD_{G_t}(v)$, then there will be no embedding $f$ of $P$ in $G$ such that $f(u)=v$, and $f(u')$ is not owned by $M_t$,  where $u' \in P$, $u' \neq u$.
\end{proposition}

Proposition~\ref{lem:diaEmbed} states that if the border distance of $v$  is not smaller than the span of query vertex $u$, there will be no cross-machine embeddings (i.e.,  embeddings where the query vertices are mapped to data vertices residing in different machines) which map $u$ to $v$. The proof of Proposition~\ref{lem:diaEmbed} is in the Appendix~\ref{subsec:spanProof}.

Let $u_{start}$ be the starting query vertex (namely, the first query vertex to be mapped) and $C(u_{start})$ be the candidate vertex set of $u_{start}$ in $G_t$.
Let $C_1(u_{start})\subseteq C(u_{start})$ be the subset of candidates whose border distance is no less than the span of $u_{start}$. According to  Proposition~\ref{lem:diaEmbed}, any embedding that maps $u_{start}$ to a vertex in $C_1(u_{start})$ can be found using a single-machine subgraph enumeration algorithm over $G_t$, independent of other machines. In \Pads, the candidates in $C_1(u_{start})$ will be processed by \sme, and the other candidates will be processed by the subsequent distributed process. 
{The \sme process is simple, and we will next focus on the distributed process.  For presentation simplicity, from now on when we say a candidate vertex of $u_{start}$, we mean a candidate vertex in $C(u_{start})-C_1(u_{start})$, unless explicitly stated otherwise.}


\ignore{  Note in \sme, we will first compute a query execution plan whose pivot vertex of first decomposition unit will be used as the starting query vertex $u_{start}$.  The details of execution plan will be discussed in Section~\ref{sec:executionPlan}.}

The distributed process consists of some daemon threads and the subgraph enumeration thread:

\item \textbf{Daemon Threads} listen to requests from other machines and support four functionalities: \\
\textit{(1) verifyE} is to return the {\em edge verification} results for a given request consisting of vertex pairs. For example, given a request $\{(v_0, v_1)$, $(v_2, v_3)\}$  posted to $M_1$, $M_1$ will return $\{true, false\}$ if $(v_0, v_1)$ is an edge in $G_1$ while $(v_2, v_3)$ is not. \\
\textit{(2) fetchV} is to return the adjacency-lists of the requested vertices of the data graph. The requested vertices sent to machine $M_i$ must reside in $M_i$. \\
\textit{(3) checkR} is to return the number of unprocessed {\em region groups} (which is a group of candidate data vertices of the starting query vertex, see Section~\ref{subsec:rmeef}) of the local machine (i.e., the machine on which the thread is running). \\
\textit{(4) shareR} is to return an unprocessed region group of the local machine to the requester machine. \textit{shareR} will also mark the region group sent out as processed.

\item \textbf{R-Meef Thread}  is the core subgraph enumeration thread. \ignore{which employs a novel framework named $\Mur$ (\rmeef).} When necessary, the local $\Mur$ thread sends \textit{verifyE} requests and \textit{fetchV} requests to the Daemon threads located in other machines, and the other machines respond to these requests accordingly. 
    
Once a local machine finishes processing its own region groups, it will broadcast a \textit{checkR} request to the other machines.
Upon receiving the  numbers of unfinished region groups from other machines, it will send a \textit{shareR} request to the machine with the maximum number of unprocessed region groups. Once it receives a region group, it will process it on the local machine. \textit{checkR} and \textit{shareR} are for load balancing purposes only, and they will not be discussed further in this paper.
\end{itemize}


\subsection{The \Mur Framework}
\label{subsec:rmeef}


Before presenting the details of the $\Mur$ framework, we need the following definitions.

\begin{defn} [embedding candidate]
Given a partition $G_t$ of data graph $G$ located in machine $M_t$ and a query pattern $P$, an injective function $f_{G_t}$: $V_P$  $\rightarrow$  $V_{G}$ is called an {\em embedding candidate (EC)} of $P$  $w.r.t$ $G_t$ if for any edge $(u$, $u')$ $\in$ $E_{P}$, there exists an edge $( f_{G_t} (u)$, $f_{G_t}(u') )$ $\in$ $E_{G_t}$ provided either $f_{G_t} (u)$ $\in$ $V_{G_t}$ or $f_{G_t}(u')$ $\in$ $V_{G_t}$.
\end{defn}

We use \ecsb to denote the set of {\ec}s of $P$ $w.r.t$ $G_t$.
Note that for an \ec $f_{G_t}$ and a query vertex $u$,  $f_{G_t}(u)$ is not necessarily owned by $G_t$. That is, the adjacency-list of $f_{G_t}(u)$ may be stored in other machines. For any query edge $(u,u')$, an \ec only requires that the corresponding  data edge $( f_{G_t} (u)$, $f_{G_t}(u') )$ exists if at least one of $f_{G_t} (u)$ and $f_{G_t}(u')$  resides in $G_t$. Therefore, an \ec may not be an embedding.
Intuitively, the existence of the edge $(f_{G_t} (u), f_{G_t}(u'))$ can only be verified in $G_t$ if one of its end vertices resides in $G_t$. Otherwise the existence of the edge  cannot be verified in $M_t$, and we call such edges {\em undetermined} edges.

\begin{defn}
Given an \ec $f_{G_t}$ of query pattern $P$, for any edge $(u,u')\in E_P$, we say $(f_{G_t}(u), f_{G_t}(u'))$ is an \emph{undetermined edge} of $f_{G_t}$ if neither $f_{G_t}(u)$ nor $f_{G_t}(u')$ is in $G_t$.
\end{defn}

\begin{exmp}
Consider a partition $G_t$ of a data graph $G$ and a triangle query pattern $P$ where $V_P=\{u_0, u_1, u_2\}$. The mapping $f_{G_t}=$ $\{(u_0, v_0)$, $(u_0, v_1)$, $(u_0, v_2)\}$ is an \ec of $P$ in $G$ $w.r.t$ $G_t$ if $v_0 \in V_{G_t}$, $v_1 \in adj(v_0)$ and $v_2 \in adj(v_0)$ and neither $v_1$ nor $v_2$ resides in $G_t$. $(v_1, v_2)$ is an undetermined edge of $f_{G_t}$.
\end{exmp}

Obviously if we want to determine whether $f_{G_t}$ is actually an embedding of the query pattern, we have to verify its undetermined edges in other machines. For any undetermined edge $e$, if its two end vertices reside in two different machines, we can use either of them to verify whether $e \in E_G$ or not. To do that, we need to send a \textit{verifyE} request to one of the machines.

Note that it is possible that an undetermined edge is shared by multiple {\ec}s. To reduce network traffic, we do not send \textit{verifyE} requests once for each individual \ec, instead, we build an {\em edge verification index} (\evi) and use it to identify {\ec}s that share undetermined edges.  We assume each \ec is assigned an ID  (We will discuss how to assign such IDs and how to build \evi in Section~\ref{sec:embeddingTrie}).


\begin{defn}[edge verification index]
Given a set \ecsb of {\ec}s, the \emph{edge verification index (\evi)} of \ecsb is a key-value map  $\ei$ where
\begin{enumerate}
\item[(1)] for any tuple $(e, IDs)\in \ei$,
\begin{itemize}
\item the key $e$ is a vertex pair $(v, v')$.
\item the value $IDs$ is the set of IDs of the {\ec}s in \ecsb of which $e$ is an undetermined edge.
\end{itemize}
\item[(2)] for any undetermined edge $e$ of $f_{G_t} \in$ \ecsb, there exists a unique tuple in $\ei$ with $e$ as the key and the ID of $f_{G_t}$ in the value.
\end{enumerate}
\end{defn}

Intuitively, the \evi groups the {\ec}s that share each undetermined edge together.  It is straightforward to see:

\begin{proposition}
Given data graph $G$, query pattern $P$ and an edge verification index $\ei$, for any $(e, IDs) \in \ei$, if $e \notin E_{G}$, then none of the {\ec}s corresponding to $IDs$ can be an embedding of $P$ in $G$.
\end{proposition}

\begin{exmp}
Consider two embedding candidates 
$f_{G_t}=$ $\{(u_0, v_0)$, $(u_0, v_1)$, $(u_0, v_2)\}$ and $f_{G_t}'=$ $\{(u_0, v_3)$, $(u_0, v_1)$, $(u_0, v_2)\}$ of a triangle pattern $P$ of a data graph $G$ where $V_P=\{u_0, u_1, u_2\}$. Assuming $(v_1, v_2)$ is an undetermined edge, we can have an edge verification index: $\ei=\{(v_1, v_2) <f_{G_t}, f_{G_t}'>\}$ where $f_{G_t}, f_{G_t}$ are represented by their IDs in $\ei$. If $(v_1, v_2)$ is verified non-existing, both $f_{G_t}$ and $f_{G_t}$ can be filtered out.

\end{exmp}

Like SEED and Twintwig, we decompose the pattern graph into small decomposition units.

\begin{defn}[decomposition]
\label{defn:decomposition}
A decomposition of query pattern $P$ is a sequence of decomposition units $\decomp$ $=$ $(dp_0$, $\dots$, $dp_l)$ where every $dp_i \in $ $\decomp$ is a subgraph of $P$ such that

\begin{enumerate}

\item[(1)] The vertex set of $dp_i$ consists of a {\em pivot} vertex $\piv$ and a non-empty set $\lf$ of {\em leaf\footnote{In an abuse of the word ``leaf".} vertices}, all of which are vertices in $V_{P}$; and for every $u' \in \lf$, $(\piv, u') \in E_{P}$.

\item[(2)] The edge set of $dp_i$ consists of two parts,  $E_{dp_i}^{star}$  and $E_{dp_i}^{sib}$, where  $E_{dp_i}^{star}$ $=$ $\bigcup_{u' \in \lf}$ $\{(dp_i.\piv,$ $u')\}$
    is the set of edges between the pivot vertex and the leaf vertices, and
    $E_{dp_i}^{sib}$ $=$ $\bigcup_{u, u' \in dp_i.\lf}$ $\{(u, u')$ $\in$ $E_{P} \}$ is the set of edges between the leaf vertices.

\item[(3)] $\bigcup_{dp_i \in \decomp}(V_{dp_i}) = V_{P}$, and for $i<j$, $V_{dp_i}\cap dp_j.\lf=\emptyset$.
\end{enumerate}
\end{defn}

Note condition (3) in the above definition says the leaf vertices of each decomposition unit do not appear in the previous units.  Unlike the decompositions in SEED \cite{DBLP:journals/pvldb/LaiQLZC16} and TwinTwig \cite{DBLP:journals/pvldb/LaiQLC15}, our decomposition unit is not restricted to stars and cliques, and $\bigcup_{dp_i \in \decomp}(E_{dp_i})$ may be a proper subset of $E_{P}$.

\begin{figure}[htbp!]
\small
\centering
\includegraphics[width=0.45\textwidth]{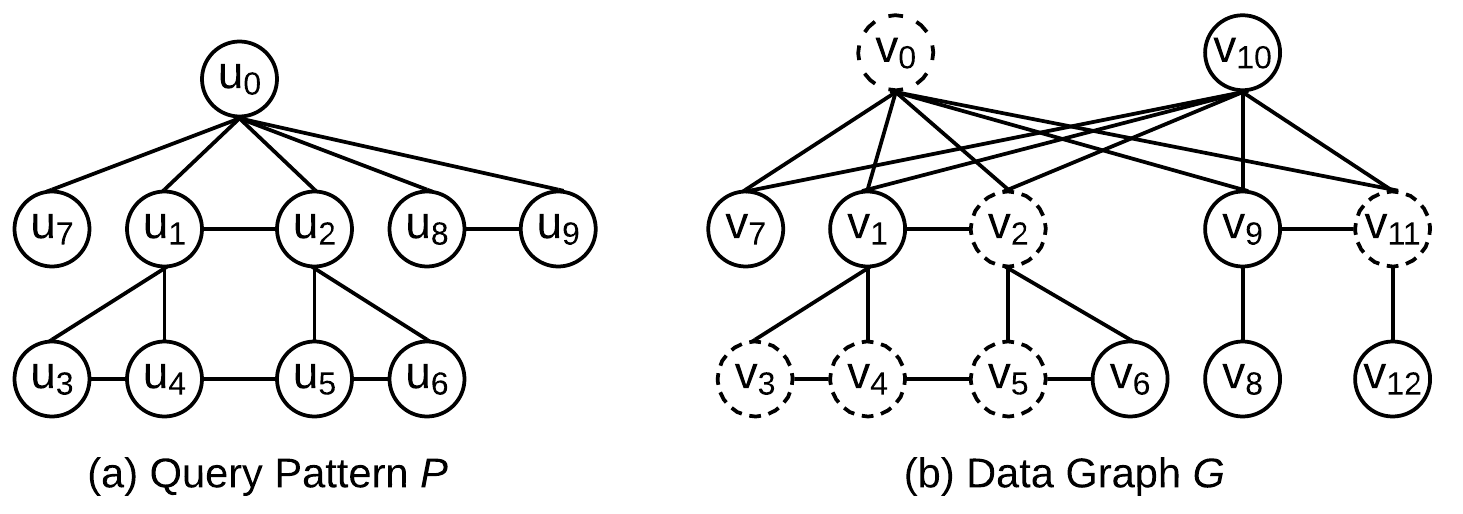}
\caption{Running Example}
\label{fig:runningExample1}
\end{figure}

\begin{exmp}
\label{exmp:executionPlan}
Consider the query pattern in Figure~\ref{fig:runningExample1}~(a),
we may have a decomposition $(dp_0$, $dp_1$, $dp_2$, $dp_3$) where $dp_0.\piv$ $=$  $u_0$, $dp_0.\lf$ $=$ $\{u_1$, $u_2$, $u_7$ $\}$, $dp_1.\piv$ $=$ $u_1$, $dp_1.\lf$ $=$ $\{u_3, u_4\}$, $dp_2.\piv$ $=$ $u_2$, $dp_2.\lf$ $=$ $\{u_5$, $u_6\}$, and $dp_3.piv=u_0$, $dp_3.\lf=\{u_8,u_9\}$. Note that the edge $(u_4$, $u_5)$ is not in any decomposition unit.
\end{exmp}

Given a decomposition $\decomp=(dp_0$,  $\dots$,  $dp_{l})$ of  pattern $P$, we define a sequence of \textit{sub-query} patterns $P_0, \ldots, P_l$, where $P_0$ = $dp_0$, and for $i>0$, $P_i$ consists of the union of $P_{i-1}$ and $dp_i$ together with the edges across the vertices of $P_{i-1}$ and $dp_i$, that is, $V_{P_i}$ = $\bigcup_{j \leq i} V_{dp_j}$, $E_{P_i}$ = $\bigcup_{j \leq i} $ $E_{P_j}$ $\cup$ $\{(u_i, u_j)\in E_P | u_i\in P_{i-1}, u_j\in dp_i.\lf\}$. Note that (a) none of the leaf vertices of $dp_i$ can be in $P_{i-1}$; and (b) $P_i$ is the subgraph of $P$ induced by the vertex set $V_{P_i}$, and $P_l =P$. We say $\decomp$ forms an {\em execution plan} if for every $i\in [1,l]$, the pivot vertex of $dp_i$ is in $P_{i-1}$\ignore{, and none of the leaf vertices of $dp_i$ is in $P_{i-1}$}. Formally, we have

\begin{defn}[execution plan]
\label{defn:executionPlan}
A decomposition $\decomp=(dp_0$,  $\dots$,  $dp_{l})$ of $P$ is  an {\em execution plan} ($\joinplan$) if $dp_i.\piv$ $\in$ $V_{P_{i-1}}$ for all $i\in [1,l]$.
\end{defn}

For example, the decomposition in Example~\ref{exmp:executionPlan} is an execution plan.

Let  $\joinplan=(dp_0$,  $\dots$,  $dp_{l})$ be an execution plan. For each $dp_i$, we define
\[ E_{dp_i}^{cro} = \{(u_i, u_j)\in E_P | u_i\in P_{i-1}, u_j\in dp_i.\lf\} (\mbox{for } i> 0)\]

We call the edges in  $E_{dp_i}^{star}$, $E_{dp_i}^{sib}$ and  $E_{dp_i}^{cro}$ the {\em expansion} edges, {\em sibling } edges, and {\em cross-unit} edges respectively. The sibling edges and cross-unit edges are both called {\em verification} edges.  

Consider $dp_0$ in Example~\ref{exmp:executionPlan}, we have $E_{dp_0}^{sib}$=$\{(u_1, u_2)\}$, $E_{dp_0}^{cro}$=$\emptyset$. For $dp_2$, we have $E_{dp_2}^{sib}$=$\{(u_5, u_6)\}$, $E_{dp_2}^{cro}$=$\{(u_4, u_5)\}$.

Note that the  expansion edges of all the units form a spanning tree of $\sPattern$, and the verification  edges are the edges not in the spanning tree.

With the above concepts, we are ready to present the $\Mur$
framework. 
Given query pattern $P$,
data graph $G$ and its partition
$G_t$ on machine $M_t$, $\Mur$
finds a set of embeddings of $\sPattern$ in
$G_t$ according to an execution plan $\joinplan$,  
%
which provides a processing order for the query
pattern $\sPattern$. In our approach, each machine $M_t$ will evaluate $P_0$ in the first round, and based on the results in round \textit{i}, it will evaluate the
next pattern $P_{i+1}$ in the next round. The final results will be obtained
when $P_{l}$ is evaluated in all machines (each machine computes a
subset of the final embeddings, the union of which is the final set
of embeddings of $\sPattern$ in $G$).

Moreover, in our approach, each machine $M_t$ starts by mapping
$dp_0.\piv$ (which is the $u_{start}$ in Section \ref{subsec:architecture}) to a candidate vertex of $dp_0.\piv$ that resides in $M_t$. 
%
When the number of such candidate vertices  is large, there is a
possibility of generating too many intermediate results (i.e., \ec{s}
and embeddings of $P_0$, $\dots$ $P_l$). To prevent memory crash, we divide
the candidate vertex set of $dp_0.\piv$ into disjoint  {\em region groups}
$\regiongroup$ = $\{rg_0$, $\dots$, $rg_h\}$, and process each group separately.

{ The workflow of \Mur is as follows:}

\begin{enumerate}[leftmargin=*]

\item[(1)] From the vertices residing in $M_t$ , $\Mur$ divides the candidate vertices of $dp_0.\piv$ into different region groups. Then it processes each group sequentially and separately.

\item[(2)] For each region group, $\Mur$ processes one unit at a round based on the execution plan $\joinplan$. In the $i^{th}$ round, the workflow can be illustrated in Figure~\ref{fig:embedFlow}.

\begin{figure}[htbp!]
\small
\centering
\includegraphics[width=0.3\textwidth]{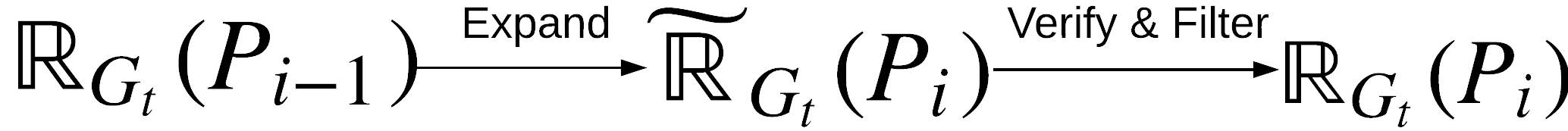}
\caption{$\Mur$ workflow}
\label{fig:embedFlow}
\end{figure}

In Figure~\ref{fig:embedFlow},  $\mathbb{R}_{G_t}(P_{i-1})$ represents the set of embedding of $P_{i-1}$ generated and cached from the last round. For the first round (i.e., round 0), $\mathbb{R}_{G_t}(P_{i-1})$ will be initialized as $\bigcup\{(dp_0.\piv, v)\}$ where $v$ is a candidate vertex of $dp_0.\piv$. By expanding $\mathbb{R}_{G_t}(P_{i-1})$, we get all the {\ec}s  of $P_{i}$ $w.r.t$ $M_t$, i.e., $\widetilde{\mathbb{R}}_{G_t}(P_{i})$. After verification and filtering, we get all the embeddings of $P_i$ for this region group of $M_t$.

In each round, the \emph{expand} and \emph{verify \& filter}  processes work as follows:

\begin{itemize}[leftmargin=*]
\item \underline{\textit{Expand}}
Given an embedding $f$ of $P_{i-1}$ obtained from the previous round, $dp_i.\piv$ has already been matched to a data vertex $v$ by $f$ since $dp_i.\piv \in P_{i-1}$. By searching the neighborhood of $v$, we expand $f$ to find  the \ec{s} of $P_i$ containing $(dp_i.\piv$, $v)$ $w.r.t$ $M_t$.  It is worth noting that if $v$ does not reside in $M_t$, we have to fetch its adjacency-list from other machines. Different embeddings from previous round may share some common foreign vertices to fetch in order to expand. To reduce network traffic, for all the embeddings from last round, we gather all the vertices that need to be fetched and then fetch their adjacency-lists together by sending a single $fetchV$ request.

One important assumption here is that each machine has a record of the ownership information (i.e., which machine a data vertex resides in) of all the vertices. This record can be constructed offline as a map whose size is $|V|$, which can be saved together with the adjacency-list and takes one extra byte space for each vertex.

\item \underline{\textit{Verify $\&$ Filter}}
Upon having a set of {\ec}s (i.e. $\widetilde{\mathbb{R}}_{G_t}(P_{i})$), we store them compactly in a {\em embedding trie} and build an \evi from them (the embedding trie and \evi will be further discussed in Section~\ref{sec:embeddingTrie}). Then we send a $verifyE$ request consisting of the keys of \evi, i.e., undetermined data edges, to other machines to verify their existence.  After we get the verification results, each failed key indicates that the corresponding {\ec}s can be filtered out. The output of the final round is the set of embeddings of query pattern $P$ found by $M_t$ for this region group.

\end{itemize}


Note that a detailed implementation and example of \Mur is given in Appendix~\ref{sec:rmeefImplementaion}. Although the idea of our framework is straightforward. However, in order to achieve the best performance, each critical component of it should be carefully designed. In the following sections, we tackle the challenges one by one.

\end{enumerate}

\section{Computing Execution Plan}
\label{sec:executionPlan}

It is obvious that we may have multiple valid execution plans for a query pattern and different execution plans may have different performance. The challenge is how to find the most efficient one among them ? In this section, we present some heuristics to find a good execution plan. 

\subsection{Minimizing Number of Rounds}
Given query pattern $P$ and an execution plan $\joinplan$, we have { $|\joinplan|+1$} rounds for each region group, and once all the rounds are processed we will get the set of final embeddings. Also, within each round, the workload can be shared. To be specific, a single undetermined edge $e$ may be shared by multiple {\ec}s.
If these embedding candidates are generated in the same round, the verification of $e$ can be shared by all of them. The same applies to the foreign vertices where the cost of fetching and memory space can be shared among multiple embedding candidates if they happen to be in the same round. Therefore, our first heuristic is to minimize the number of rounds (namely, the number of decomposition units) so as to maximize the workload sharing. 

Here we present a technique to compute a query execution plan, which guarantees a minimum number of rounds. Our technique is based on the concept of {\em maximum leaf spanning tree} \cite{fernau2011exact}.

\begin{defn}
A {\em maximum leaf spanning tree (\mlst)} of pattern $\sPattern$ is a spanning tree of $\sPattern$ with the maximum number of leafs (a leaf is a vertex with degree 1).
The number of leafs in a \mlst of $\sPattern$ is called the {\em maximum leaf number} of $\sPattern$, denoted $l_{\sPattern}$.
\end{defn}

A closely related concept is {\em minimum connected dominating set}.

\begin{defn}
A {\em connected dominating set (\cds)}  of $\sPattern$ is a subset $D$ of $V_{\sPattern}$  such that (1) $D$ is a dominating set of $\sPattern$, that is,
any vertex of $\sPattern$ is either in $D$ or adjacent to a vertex in $D$, and (2) the subgraph of $\sPattern$ induced by $D$ is connected.

A {\em minimum connected dominating set (\mcds)} is  a \cds with the smallest cardinality among all \cds{s}.  The number of vertices in a  \mcds is called the {\em connected domination number}, denoted $c_{\sPattern}$.
\end{defn}

It is shown in \cite{MLST92} that \ignore{the set of non-leaf vertices of a \mlst of query $P_{\sPattern}$ is a \mcds  of $P$, therefore} $|V_{\sPattern}|=c_{\sPattern}+l_{\sPattern}$.

\begin{theorem}                                                                                                                                                                                                                         Given a  pattern $\sPattern$, any execution plan of $\sPattern$ has at least $c_{\sPattern}$ decomposition units, and there exists an execution plan with exactly $c_{\sPattern}$ decomposition units.
\label{the:miniPlan}
\end{theorem}

The proof of Theorem~\ref{the:miniPlan} is in the Appendix~\ref{subsec:spanProof}. 

Theorem~\ref{the:miniPlan} indicates that $c_{\sPattern}$ is the minimum number of rounds of any execution plan. {The above proof  provides a method to construct an execution plan with $c_{\sPattern}$ rounds from a \mlst.} It is worth noting that the decomposition units in the query plan constructed as in the proof have distinct pivot vertices.

\begin{exmp}
Consider the pattern $\sPattern$, 
it can be easily verified that the tree obtained by erasing the edges $(u_1,u_2)$,  $(u_3, u_4)$, $(u_4,u_5)$, $(u_5,u_6)$ and $(u_8, u_9)$ is a \mlst of $\sPattern$. Choosing $u_0$ as the root, we will get a minimum round execution plan $\joinplan_1$=$\{dp_0$, $dp_1$, $dp_2\}$ where $dp_0.\piv$ $=$ $u_0$, $dp_0.\lf$ $=$ $\{u_1$, $u_2$, $u_7$, $u_8$, $u_9\}$, $dp_1.\piv$ $=$ $u_1$, $dp_1.\lf$ $=$ $\{u_3, u_4\}$ and $dp_2.\piv$ $=$ $u_2$, $dp_2.\lf$ $=$ $\{u_5, u_6\}$.
If we choose $u_1$ as the root, we will get a different minimum-round execution plan $\joinplan_2$=$\{dp_0, dp_1, dp_2 \}$, where
$dp_0.\piv=u_1$, $dp_0.\lf=\{u_0,u_3, u_4\}$,
$dp_1.\piv=u_0$, $dp_1.\lf=\{u_2$, $u_7$ $u_8,u_9\}$,
$dp_2.\piv = u_2$, $dp_2.\lf=\{u_5, u_6\}$

\label{example4}
\end{exmp}

\subsection{Minimizing the span of {$dp_0.piv$}}
Given a pattern $\sPattern$, multiple execution plans may exist with the minimum number of rounds, while their $dp_0.piv$ can be different. When facing this case, here we present our second heuristic which is to choose the plan(s) whose $dp_0.piv$ have the smallest span. This strategy will maximize the number of embeddings that can be found using \sme.
Recall the \Pads architecture where $dp_0.piv$ is the starting query vertex $u_{start}$, based on Proposition~\ref{lem:diaEmbed}, we know that the more candidate vertices of $dp_0.piv$ can be processed in \sme, the more workload can be separated from the distributed processing, and therefore the more communication cost and memory usage can be reduced.

\begin{figure}[htbp!]
\small
\centering
\includegraphics[width=0.2\textwidth]{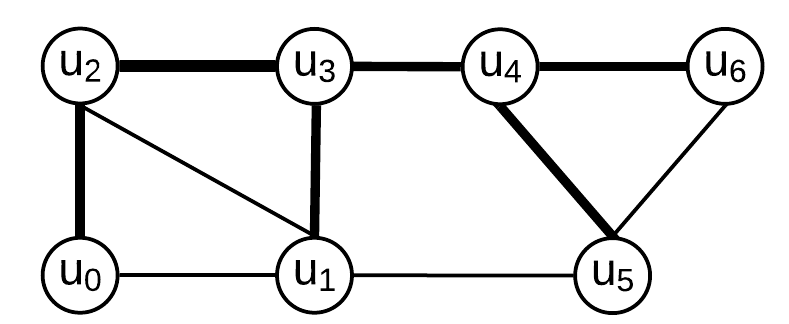}
\caption{A Query Pattern}
\label{fig:spanPattern}
\end{figure}

Consider the pattern in Figure~\ref{fig:spanPattern}, the bold edges demonstrate a  \mlst based on which both $u_3$ and $u_4$ can be chosen as  $dp_0.\piv$. And the execution plans from them have the same number of rounds.  However, $\QD_{P}(u_3)=2$ while $\QD_{P}(u_4)=3$. Therefore we choose the plan with $u_3$ as the $dp_0.\piv$.

\subsection{Maximizing Filtering Power}
\label{subsec:movingFV}
Given a pattern $\sPattern$, multiple execution plans may exist with the minimum number of rounds and their $dp_0.piv$ have the same smallest span.
Here we use the third heuristic which is to choose plans with more verification edges in the earlier rounds. The intuition is to maximize the filtering power of the verification edges as early as possible.
To this end, we propose the following score function $\mathcal{SC}(\joinplan)$ for an execution plan $\joinplan=\{dp_0$,  $\dots$,  $dp_{l}\}$:

\begin{equation}
\label{equ:scorePlan}
\mathcal{SC}(\joinplan) = \sum_{dp_i \in \joinplan} \frac{1}{(i + 1)^\rho} \times (|E_{dp_i}^{sib}| + |E_{dp_i}^{cro}|)
\end{equation}
$|E_{dp_i}^{sib}| + |E_{dp_i}^{cro}|$ is the number of verification edges in round $i$, and $\rho$ is a positive parameter used to tune the score function. In our experiments we use $\rho =1$. The function $\mathcal{SC}(\joinplan)$ calculates a score by assigning larger weights to the verification edges in earlier rounds (since  $\frac{1}{(i + 1)^\rho} > \frac{1}{(j + 1)^\rho}$ if $i<j$).

\begin{exmp}
Consider the query plans $\joinplan_1$ and $\joinplan_2$ in Example~\ref{example4}. The total number of verification edges in these plans are the same. In $\joinplan_1$, the number of verifications edges for the first, second and third round is 2, 1, 2 respectively.  In $\joinplan_2$, the number of verification edges for the three rounds is 1, 2, and 2 respectively. Therefore, we prefer $\joinplan_1$. Using $\rho=1$, we can calculate the scores of the two plans as follows:

$\mathcal{SC}(\joinplan_1)=$ $2/1+ 1/2 + 2/3$ $\approx 3.2$

$\mathcal{SC}(\joinplan_2)=$ $1/1+ 2/2 + 2/3$ $\approx 2.7$

\label{example5}
\end{exmp}

When several minimum-round execution plans have the same score,  we use another heuristic rule to choose the best one from them: the larger the degree of the pivot vertex, the earlier we process the unit. The pivot vertex with a larger degree has a stronger power to filter unpromising candidates.

To accommodate this rule, we can modify the score function in (1) by adding another component as follows:
\begin{equation}
\label{equ:scorePlan1}
\mathcal{SC}(\joinplan)=\sum_{dp_i \in \joinplan}[\frac{|E_{dp_i}^{sib}| + |E_{dp_i}^{cro}|}{(i + 1)^\rho} + \frac{deg(dp_i.\piv)}{(i + 1)}]
\end{equation}

To this end, we have a set of rules to follow when to compute the execution plan. Since the query vertex is normally very small. We can simply enumerate all the possible execution plans and choose the best according to those rules.

\section{Embedding Trie}
\label{sec:embeddingTrie}
As stated before, to save memory, the intermediate results (which include embeddings and embedding candidates generated in each round) are stored a compact data structure called an embedding trie.  Besides the compression, the challenges here are how to ensure each intermediate result has a unique ID in the embedding trie and the embedding trie can be easily maintained ? 



Before we give our solution, we first define a matching order, which is the order following which the query vertices are matched in \Mur. It is also the order the nodes in the \et are organized.
{
\begin{defn}[Matching Order]
  Given a query execution plan  $\joinplan$ $=$ $\{dp_0$, $\ldots$, $dp_l\}$ of pattern  $\sPattern$, the matching order w. r. t $\joinplan$ is a relation  $\prec$  defined over the vertices of $P$ that satisfies the following conditions:
   \begin{enumerate}
   \item $dp_i.piv \prec$ $dp_j.piv$ if $i<j$;
   \item For any two vertices $u_1\in dp_i.\lf$ and $u_2\in dp_{j}.\lf$, $u_1\prec u_2$ if $i<j$.

   \item For  $i\in[0,l]$:
	\begin{enumerate}[label = (\roman*)]
	\item $dp_i.\piv\prec u$ for all $u\in dp_i.\lf$;
	\item for  any vertices $u_1$, $u_2\in dp_i.\lf$ that are not the pivot vertices of other units,  $u_1\prec u_2$ if $deg(u_1)$ > $deg(u_2)$, or $deg(u_1) = deg(u_2)$ and the vertex ID of $u_1$ is less than that of $u_2$;
	\item if $u_1\in dp_i.\lf$ is a pivot vertex of another unit, and $u_2\in dp_i.\lf$ is not a pivot vertex of another unit, then $u_1\prec u_2$.
    \end{enumerate}	
   \end{enumerate}
\label{matchorder}
\end{defn}

Intuitively the above relation orders the vertices of $P$ as follows:
(a) Generally a vertex $u_1$ in $dp_i$ is before a vertex $u_2$ in $dp_j$ if $i<j$, except for the special case where $u_1\in dp_i.\lf$ and $u_2=dp_j.piv$. In this special case, $u_2$ may appear in the leaf of some previous unit $dp_k$ ($k\leq i$), and it may be arranged before $u_1$ according to Condition (2) or Condition (3) (ii).
\ignore{$dp_j.\piv$ if $dp_j.\piv$ is a leaf vertex of some $dp_k$ ($k\leq i$), in which case its position has been determined when the leaves of $dp_k$ are arranged.}
(b) Starting from $dp_0$, the vertex $dp_0.\piv$ is arranged before all other vertices. For the leaf vertices of $dp_0$, it arranges those that are pivot vertices of other units before those that are not (Condition (3)(iii)), and for the former, it arranges them according to the ID of the units for which they are the pivot vertex\footnote{Note that no two units share the same pivot vertex.} (Condition (1)); for the latter, it arranges them in descending order of their degree in the original pattern $P$, and if they have the same degree it arranges them in the order of vertex ID (Condition (3) (ii)).
For each subsequent $dp_i$, the pivot vertex must appear in the leaf of some previous unit, hence its position has been fixed; and the leaf vertices of $dp_i$ are arranged in the same way as the leaf vertices of $dp_0$.
}

It is easy to verify $\prec$ is a strict total order over $V_P$. Following the matching order, the vertices of $P$ can be arranged into an ordered list.
Consider the execution plan $\joinplan_1$ in Example~\ref{example4}. The vertices in the query can be arranged as ($u_0, u_1,u_2, u_7, u_8,u_9,u_3,u_4,u_5,u_6$) according to the matching order.


Let $\joinplan=\{dp_0$, $\ldots$, $dp_l\}$ be an execution plan,  $P_i$ be the subgraph of $P$ induced from the vertices in $dp_0\cup \cdots \cup dp_i$ (as defined in Section~\ref{subsec:rmeef}), and $\mathbb{R}$ be a set of {\em results} (i.e., embeddings or embedding candidates) of $P_i$.  For easy presentation, we assume the vertices in $P_i$ have been arranged into the list $u^0, u^1, \ldots, u^n$ by the matching order, that is, the query vertex at position $j$ is $u^j$. Then each result of $P_i$ can be represented as a list of corresponding data vertices. These lists can be merged into a collection of trees as follows:

\begin{enumerate}
\item[(1)] Initially, each result $f$ is treated as a tree $T_f$, where the node at level $j$ stores the data vertex $f(u^j)$ for $j\in [0,n]$, and the root is the node at level 0.

\item[(2)] If multiple results map $u^0$ to the same data vertex, merge the root nodes of their trees. This partitions the results in $\mathbb{R}$ into different groups, each group will be stored in a distinct tree.


\item[(3)] For each newly merged node $\trienode$, if multiple children of  $\trienode$ correspond to the same vertex, merge these children into a single child of $\trienode$.

\item[(4)] Repeat step (3) untill no nodes can be merged.
\end{enumerate}

The collection of trees obtained above is a compact representation of the results in $\mathbb{R}$. Each leaf node in the tree uniquely identifies a result.

The embedding trie is a collection of similar trees. However, since the purpose of the embedding trie is to save space, we cannot get it by merging the result lists. Instead, we will have to construct it by inserting nodes one by one when results are generated, and removing nodes when results are eliminated. Next we formally define \et  and present the algorithms for the maintenance of the \et.

\subsection{Structure of the Embedding Trie}

\begin{defn}[Embedding Trie]
 Given a set $\mathbb{R}$ of results of $P_{i}$, the {\em embedding trie} of $\mathbb{R}$ is a collection of trees used to store the results in $\mathbb{R}$ such that:
\begin{enumerate}
 \item Each tree represents a set of results that map $u^0$ to the same data vertex.

 \item Each tree node $\trienode$ has
    \begin{itemize}
    \item \textbf{v}:  a data vertex
    \item \textbf{parentN}:   a pointer pointing to its parent node (the pointer of the root node is null).

    \item \textbf{childCount}: the number of child nodes of $\trienode$.
    \end{itemize}

\item If two nodes have the same parent, then they store different data vertices.

\item Every leaf-to-root path represents a result in $\mathbb{R}$, and every result in $\mathbb{R}$ is represented as a unique leaf-to-root path.

\item If we divide the tree nodes into different levels such that the root nodes are at level 0, the children of the root nodes are at level 1 and so on, then the tree nodes at level $j$ ($j\in [0,1]$) store the set of values $\{f(u^j) | f\in$ $\mathbb{R}$ $\}$.
\end{enumerate}

\end{defn}


%
%
%
%

\begin{figure}[htbp!]
\small
\centering
\includegraphics[width=0.45\textwidth]{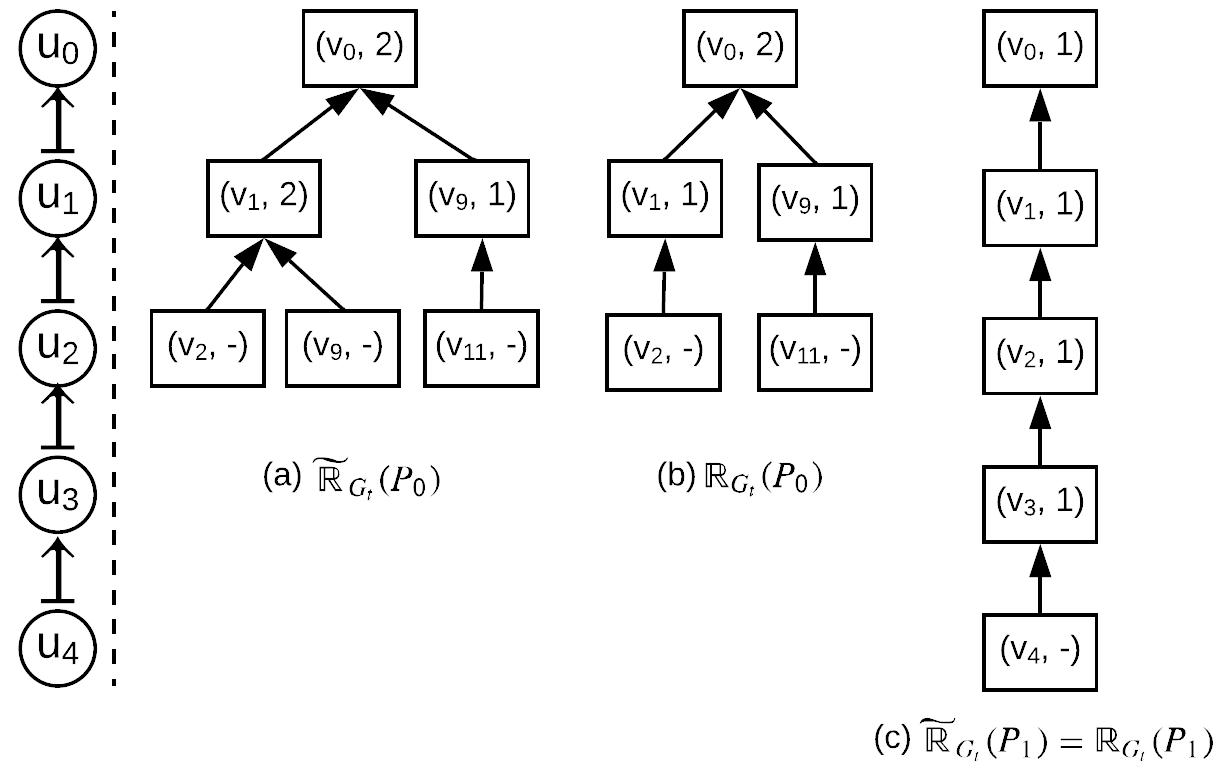}
\caption{Example of Embedding Trie}
\label{fig:embTrie}
\end{figure}


\begin{exmp}
Consider $P_0$ in Example~\ref{exmp:runningExmp}, where the vertices are ordered as $u_0,u_1, u_2$ according to the matching order. There are three \ec{s} of $P_0$: $(v_0,v_1,v_2)$, $(v_0,v_1,v_9)$ and $(v_0,v_9,v_{11})$. These results can be stored in a tree shown in   Figure~\ref{fig:embTrie}(a).
When the second \ec is filtered out, we have $\mathbb{R}_{G_t}(P_{0})$ compressed in a tree as shown in Figure~\ref{fig:embTrie}(b).
The first \ec can be expanded to an \ec of $P_1$ (where the list of vertices of $P_1$ are $u_0, u_1,u_2,u_3,u_4$), which is as shown in Figure~\ref{fig:embTrie}(c).
\end{exmp}


\ignore{
\begin{algorithm}
\DontPrintSemicolon
\small
\SetAlgoLined
\KwIn{A failed trie node $\trienode$}
$\trienode' \gets \trienode.childN$ \\
\While{$\trienode$ is not Null }
{
		$\trienode'.parentNodeNo--$ \\
		\If{$\trienode'.parentNodeNo$ $= 0$}
		{
			$\trienode'' \gets  \trienode'$
			$\trienode'  \gets \trienode'.childN$       \\
			remove $\trienode''$ from $\invTrie$ \\ 					
		}
}
remove $\trienode$ from $\invTrie$ \\
\caption{\sc removeTrieNode}
\label{algorithm:removeTrieNode}
\end{algorithm}
}

Although the structure of embedding trie is simple,  it has some nice properties:

\begin{itemize}[leftmargin=*]
\item \underline{\textit{Compression}} Storing the results in the \et saves space than storing them as a collection of lists.


\item \underline{\textit{Unique ID}} For each result in the embedding trie, the  address of its \tNode in memory can be used as the unique ID.

\item \underline{\textit{Retrieval}} Given a particular ID represented by a \tNode, we can easily follow its pointer $parentN$ step-by-step to retrieve the corresponding result.

\item \underline{\textit{Removal}}
To remove a result with a particular ID, we can remove its corresponding \tNode and decrease the $childCount$ of its parent node by 1. If $ChildCount$ of this parent node reaches 0, we remove this parent node. This process recursively affects the ancestors of the \tNode.

\end{itemize}

\subsection{Maintaining the \et}

Recall that in Algorithm~\ref{algorithm:disSubEnum}, given an embedding $f$ of $P_{i-1}$,  the function $expandEmbedTrie$ is used to search for the \ec{s} of $dp_i$ within the neighbourhood of the mapped data vertex of $v_{piv}$, where $v_{piv}=f(dp_i.\piv)$. Moreover, the $expandEmbedTrie$ function handles the task of expanding the {\et} {$\eTrie$} by concatenating $f$ with each newly found \ec of $dp_i$. If an \ec is filtered out or if an embedding cannot be expanded to a final result, the function must remove it from $\eTrie$. Now we present the details of the $expandEmbedTrie$ function in Algorithm~\ref{algorithm:expandEmbedTrie}.

When $dp_i.\piv$ is mapped to the data vertex $v=f(dp_i.\piv)$ by an embedding $f$ of $P_{i-1}$, Algorithm~\ref{algorithm:expandEmbedTrie} uses a backtracking approach to find the \ec{s} of $P_i$  within the neighbourhood of $v$. The recursive procedure is given in the subroutine $adjEnum$. In each round of the recursive call, $adjEnum$ tries to match $u'$ to a candidate vertex $v$ and add $(u'$,$v)$ to $f$, where $u'$ is a query vertex in $dp_i.\lf$. When $f$ is expanded to an \ec of $P_i$, which means an \ec of $dp_i$ is concatenated to the original $f$, we add it into $\eTrie$ by chaining up the corresponding \et nodes. If $f$ cannot be expanded into an \ec of $P_i$, we will remove it from $\eTrie$.

\begin{algorithm}
\DontPrintSemicolon
\KwIn{an embedding $f$ of $P_{i-1}$, local machine $M_t$, unit $dp_i$, \et $\eTrie$}
\KwOut{expanded $\eTrie$ and an edge verification index \ei}

 $v_{} \gets f(dp_i.\piv)$ \\
 \For {\textbf{each} $ u \in dp_i.\lf$ }
		{
		$C(u) \gets adj(v_{})$ \\
	  	\For{each $(u, u')$ $\in$ $E_{dp_i}^{cro}$}
			{
			\If{$f(u')$ resides in $M_t$}
				{
				$C(u) \gets adj(f(u')) \cap C(u) $\\
				}
	 		}
		\If{$C(u)=\emptyset$}
			{
				remove $f$ from $\eTrie$ \\			
				\Return
			}
		}
				
  $u$ $\gets$ next vertex in query vertex list \\				
 get $\trienode$ corresponding to $f$ 	   \\
 $\sc adjEnum$($\trienode$, $u$)    \\

\caption{\sc expandEmbedTrie}
\label{algorithm:expandEmbedTrie}
\end{algorithm}

Lines 1 to 9 of Algorithm~\ref{algorithm:expandEmbedTrie} compute the candidate set for each $u\in dp_i.\lf$ as the intersection of the neighbor set of $v=f(dp_i.\piv)$ and the neighbor set of each $f(u')$, where $(u', u)$ is a cross-unit edge and $f(u')$ is in $M_t$. {If any of the candidate sets is empty, it removes $f$ from $\eTrie$. Otherwise} it passes on the next query vertex $u$ and the ID of $f$ (which is a node in $\eTrie$) to the recursive subroutine $adjEnum$.


The subroutine $adjEnum$ is given in Algorithm~\ref{algorithm:adjEnum}. It plays the same roles as the {\em SubgraphSearch} procedure in the backtracking framework \cite{DBLP:journals/pvldb/LeeHKL12}.
%
 In Line 1, $adjEnum$ creates an local variable $\mathcal{F}_{current}$ with default value $false$. The value indicates whether $f$ can be extended to an \ec of $P_i$.
For the leaf vertex $u$, $adjEnum$ first creates a copy  $C_{r}(u)$ of $C(u)$, and then refines the candidate vertex set $C_r(u)$ by considering every sibling edge $(u,u')$ where $u'$ has already been mapped by $f$ to $f(u')$. If $f(u')$ resides in $M_t$,  $C_{r}(u)$ is shrank by an intersection with $adj(f(u'))$ (Line 2 to 5).
 Then, for each vertex $v$ in the refined set $C_{r}(u)$, it first initializes a flag $\mathcal{E}$ with the value $true$ (Line 7), this value indicates whether $u$ can be potentially mapped to $v$. Then if $v$ resides in $M_t$ it will check every verification edge $(u,u')$ where $u'$ has been mapped to see if $(v, f(u'))$ exists, if one of such edge does not exist, it will set $\mathcal{E}$ to false (Lines 8 to 11), meaning $u$ cannot be mapped to $v$. This part (Lines 7 to 11) is like the {\em IsJoinable} function in the backtracking framework \cite{DBLP:journals/pvldb/LeeHKL12}.

%

If $\mathcal{E}$ is still true after the local verification, we add $(u$, $v)$ to $f$ (Line 13). Then we create a new trie node $\trienode'$ for $v$ with $\trienode$ as its \emph{parentN} (Line 14, 15).
After that, if $f$ grows to an \ec of $P_i$ , then for each undetermined edge $e$ of $f$ (both end vertices are not in the local machine), we add $\trienode'$ to $\ei[e]$ (Line 17, 18). We also set the $\mathcal{F}_{current}$ as true (Line 19). If $f$ is not an \ec of $P_i$, which means there are still leaf vertices of $dp_i$ not matched, we get the next leaf vertex $u'$ (Line 21), and launch a recursive call of $adjEnum$ by passing it $\trienode'$ and $u'$ (Line 22). We record the return value from its deeper $adjEnum$ as $\mathcal{F}_{deeper}$.
If $\mathcal{F}_{deeper}$ is true after all the recursive calls, which means there are \ec{s} with $v$ mapped to $u$ in $f$, we increase $childCount$ of the \pNode $\trienode$ and add the newly created $\trienode'$ to as a child of $\trienode$ in $\eTrie$ (Line 23 to 25).   Then we backtrack by removing $(u, v)$ from $f$, so that we can try to map $u$ to another candidate vertex in $C_r(v)$.

After we tried all the candidate vertices of $C_r(u)$, we return the value of $\mathcal{F}_{current}$ (Line 27).

\begin{algorithm}
\DontPrintSemicolon
\KwIn{ Trie node $\trienode$ representing embedding $f$ of $P_{i-1}$ , leaf vertex $u$ of $dp_i$}
\KwOut{expanded $\eTrie$ and an edge verification index $\ei$}
$\mathcal{F}_{current}$ $\gets false$  \\
 $C_{r}(u) \gets C(u)$ \\
 \For{each $u'$ mapped in $f$ and $(u, u') \in E_{dp_i}^{sib}$}
{
		\If{$f(u')$ resides in $M_t$}
			{
			$C_{r}(u) \gets  adj(f(u')) \cap C_{r}(u)$													
			}
}
		
\For{\textit{each} $v \in C_{r}(u)$}
{
    $\mathcal{E} \gets true$ \\

    \If {$v$ resides in $M_t$}
       {
		\For{each $(u, u') \in (E_{dp_i}^{sib} \cup E_{dp_i}^{cro})$ and $u'$ mapped in $f$ }
		   {
     			\If{$(v, f(u'))$ not exists}
				{
				  $\mathcal{E}  \gets false$ \\
				}
			
		   }
		}	
		\If{$\mathcal{E}$ is true}
		{
			add $(u, v)$ to $f$ \\
			create a trie node $\trienode'$ \\
			$\trienode'.v \gets v$, $\trienode'.parentN \gets \trienode$ \\
			\If{$|f| = |V_{P_{i}}|$}
			{
				\For{each undetermined edge $e$ of $f$}
				{
					add $\trienode'$ to $\ei[e]$ \\
				}
				$\mathcal{F}_{current} \gets true$  \\
			}
			\Else
			{
					 $u'$ $\gets$ next vertex in $dp_i.\lf$ \\
					 $\mathcal{F}_{deeper} \gets \textit{adjEnum}(\trienode', u')$ \\								
			}
			
			\If{$F_{deeper}$ is true}
			{
				$\trienode.childCount++$ \\
				add $\trienode'$ as a child node of $\trienode$ in $\eTrie$\\		
			}
				
			remove $(u, v)$ from $f$ \\	
		}
}
 return $\mathcal{F}_{current}$ \\
\caption{\sc adjEnum}
\label{algorithm:adjEnum}
\end{algorithm}

Note that the edge verification index $\ei$ is maintained during the expansion process.

\section{Memory Control Strategies}
\label{sec:memoryControl}
This section focuses on the challenge of robustness of $\Mur$. Since \Mur still caches fetched foreign vertices and intermediate results in memory,  memory consumption is still a critical issue when the data graph is large. We propose a grouping strategy to keep the peak memory usage  under the memory capacity of the local machine.




Our idea is to divide the candidate vertices of the first query vertex $dp_0.\piv$ into disjoint groups and process each group independently. In this way, the overall cached data on each machine will be divided into several parts, where each part is no larger than the available memory $\Phi$.

\begin{figure}[htbp!]
\small
\centering
\includegraphics[width=0.3\textwidth]{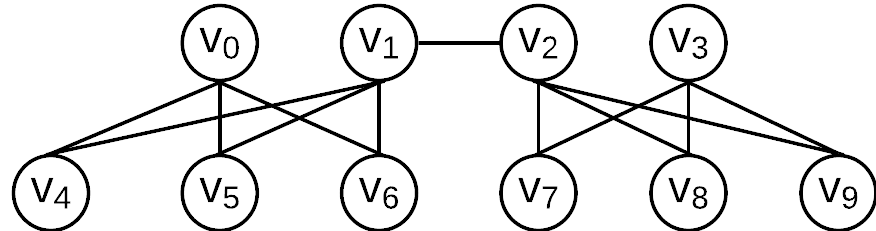}
\caption{Grouping Example}
\label{fig:regionGroup}
\end{figure}

A naive way of grouping the candidate vertices is to divide them randomly. However,  random grouping of the vertices may put vertices that are ``dissimilar" to each other into the same group, potentially resulting in more network communication cost. Consider the data graph in Figure~\ref{fig:regionGroup}. Suppose the candidate vertex set is $\{v_0$,$v_1$,$v_2$, $v_3\}$. If we divide it into two groups $\{v_0, v_1\}$ and $\{v_2,v_3\}$, then because $v_0$ and $v_1$ share most neighbours, there is a good chance for the  \ec{s} of $dp_0$ generated from $v_0$ and $v_1$ to share common verification edges\ignore{ (e.g., $(v_5,v_6)$ if $v_5$ and $v_6$ are not on the local machine)}, and share common foreign vertices that need to be fetched (e.g., if $dp_1.\piv$ is mapped to $v_5$ by \ec{s} originated from $v_0$ and $v_1$,  and $v_5$ is not on the local machine). However, if we partition the candidate set into $\{v_0,v_2\}$ and $\{v_1, v_3\}$, then there is little chance for such sharing.

Our goal is to find a way to partition the candidate vertices into groups so that the chance of  edge verification sharing and foreign vertices sharing by the results in each group is maximized.

Let $C\equiv C(dp_0.\piv)$ be the candidate set of $dp_0.\piv$, and $\Phi$ be the available memory. 
Our method is to generate the groups one by one as follows. First we pick a random vertex $v\in C$ and let  $rg=\{v\}$ be the initial group. If the estimated memory requirement of the results originated from   $rg$, denoted $\phi(rg)$ (we will discuss memory estimation shortly),  is less than $\Phi$, we choose another candidate vertex in $C-rg$ that has the greatest proximity to $rg$ and add it to $rg$; if $\phi(rg)$ > $\Phi$ we remove the last added vertex from $rg$. This generates the first group. For the remaining candidate vertices we repeat the process, until all candidate vertices are divided into groups. The detailed algorithm is given in the Algorithm~\ref{algorithm:regionGroups}.  Here an important concept is the the proximity of a vertex $v$ to a group of vertices, and we define it as the percentage of $v'$s neighbors that are also neighbors of some vertex in $rg$, that is,

\begin{equation}
proximity(v,rg) =  \frac{|adj(v)\cap \bigcup_{v'\in rg} adj(v')|}{|adj(v)|}
\end{equation}

Intuitively the vertices put into the same group are within a region - each time we will choose a new vertex that has a distance of at most 2 from one of the vertices already in the group (unless there are no such vertices). Therefore we call the group a {\em region} group.

\stitle{Estimating memory usage}
In our system, the main memory consumption comes from the intermediate results and the fetched foreign vertices. The space cost of other data structures is trivial.

Consider the set of intermediate results $\mathcal{R}$ originated from the group $rg \subseteq C$. Recall that all results originated from the same candidate vertex of $dp_0.\piv$ are stored in the same tree, while any results originated from different candidate vertices are stored in different trees. Therefore, if we know the space cost of the results originated from every candidate vertex, we can add them together  to obtain the space cost of all results originated from $rg$.

{  To estimate the space cost of the results originated from a single vertex, we use the average space cost of local embeddings of a candidate vertex $v \in C_1(u_{start})$ in \et format, which can be obtained when we conduct \sme.
Recall that for each $v$ of $C_1(u_{start})$ in \sme, we find the local embeddings originated from $v$ following a backtracking approach. In each recursive step of the backtracking approach, we may record the number of candidate vertices that are  matched to the corresponding query vertex. The sum of all steps will be the number of trie nodes if we group the those local embeddings into \et. Based on the sum, we know the space cost of local embeddings originating from $v$ in the format of \et. \ignore{Therefore we can easily obtain the average space cost of a single vertex in \sme.}
}.

Next, we consider the space cost of the fetched foreign vertices in each round. Recall that when expanding the embeddings of $P_{i-1}$ to \ec{s} of $P_i$, we only need to fetch vertex $v$ if there exists $f\in \ecsp{P_{i-1}}$ such that $f(dp_i.\piv)=v$.  In the worst case, for every candidate vertex $v$ of $dp_i.\piv$, there exists some  $f\in \ecsp{P_{i-1}}$ which maps  $dp_i.\piv$ to $v$, and none of these candidate vertices of $dp_i.\piv$  resides locally. Therefore the number of data vertices that need to be fetched equals to  $|C(dp_i.\piv)|$ in the worst case.

In practice, the space cost of  $C(dp_i.\piv)$ is usually small compared with that of the intermediate results, and we can allocate a certain amount of memory for caching the fetched data vertices. Note that when more data vertices need to be fetched, we may release some previously cached data vertices if necessary.   Therefore we can ignore the space cost of the fetched data vertices when we estimate the memory cost of each region group.


\begin{algorithm}
\DontPrintSemicolon
\small{
\SetAlgoLined
\KwIn{the candidate vertex set $C=C(dp_0.\piv)$ on $M_t$}
\KwOut{A region group $rg$}



Pick a random vertex $v\in C$ \\

$rg \gets \{v\}$ \\

$C \gets C-\{v\}$ \\

\While{$C \neq \emptyset$ $\wedge$  $\phi(rg)< \Phi$ }
{
   $v\gets $ $\operatorname*{arg\,max}_{v\in C} (proximity(v,rg))$ \\

   $rg \gets rg \cup \{v\}$ \\

   $C \gets C-\{v\}$ \\
}

\If{$\phi(rg) > \Phi$}
 { remove last added vertex from $rg$ and put it back  into $C$
 }

} 

return $rg$

\caption{\sc FindRegionGroups}
\label{algorithm:regionGroups}
\end{algorithm}

\section{Experiment}
\label{sec:experiment}
In this section, we present our experimental results.

{\stitle{Environment}}
We conducted our experiments in a cluster platform where each machine is equipped with Intel CPU with 16 Cores and 16G memory. The operating system of the cluster is Red Hat Enterprise Linux 6.5.

{\stitle{Algorithms}}
We compared our system with four state-of-the-art distributed subgraph enumeration approaches:
\begin{itemize}
\item \psgl \cite{DBLP:conf/sigmod/ShaoCCMYX14}, the algorithm using graph exploration originally based on Pregel.
\item \twintwig\cite{DBLP:journals/pvldb/LaiQLC15}, the algorithm using joining approach originally based on MapReduce.
\item \seed\cite{DBLP:journals/pvldb/LaiQLZC16},  an upgraded version of \twintwig while supporting clique decomposition unit.
\item \qiao\cite{DBLP:journals/pvldb/QiaoZC17}, the algorithm relying on clique-index and compression and originally using MapReduce.
\end{itemize}

We implemented our approach in C++ with the help of Mpich2 \cite{mpich2} and Boost library \cite{boostlibrary}. We used Boost.Asio to achieve the asynchronous message listening and passing. We used TurboIso\cite{turboIso} as our \sme processing algorithm.

The performance of distributed graph algorithms varies a lot depending on different programming languages and different underline distributed engine and file systems \cite{DBLP:journals/pvldb/AmmarO18}. It is not fair enough to simply compare our approach with the Pregel-based \psgl or  other Hadoop-based approaches. Therefore to achieve a fair comparison, we implemented \psgl, \twintwig and \seed  using C++ with MPI library. {  For \qiao, we chose to use the original program provided by its authors because our experiments with \twintwig and \seed indicate that our implementation and the original implementation over Hadoop showed no significant difference in terms of performance.} In memory, we loaded the data graph in each node in the format of adjacency-list for \Pads, \psgl and \twintwig. In order to support the clique decomposition unit of \seed, we also loaded the edges in-memory between the neighbours of a vertex along with the adjacency-list of the vertex.

{\stitle{Dataset \& Queries}} We used four real datasets in our experiments: DBLP, RoadNet, LiveJournal and UK2002.  The profiles of these data sets are given in Table~\ref{t:datasets}. The diameter in Table~\ref{t:datasets} is the longest shortest path between any two data vertices. We partitioned each data graph using the multilevel k-way partition algorithm provided by Metis \cite{metis}. DBLP is a relatively small data graph which can be loaded into memory without partitioning, however, we still partition it here. One may argue when the data graph is small, we can use single-machine enumeration algorithms. However, our purpose of using DBLP here is not to test which algorithm is better when the graph can be loaded as a whole, but is to test whether the distributed approaches can fully utilize the memory when there is enough space available.

{ RoadNet is a larger but much sparser data graph than the others,  consequently the number of embeddings of each query is smaller. Therefore it can be used to illustrate whether a subgraph enumeration solution has good filtering power to filter out false embeddings early. In contrast, the two denser data graphs, liveJournal and UK2002, are used to test the algorithms' ability to handle denser graphs with huge numbers of embeddings.}

\begin{table}[h]
    \caption{Profiles of datasets}
    \small
	\centering
	\label{t:datasets}
 	\begin{tabular}{|c|c|c|c|c|}
    \hline
     \textbf{Dataset($G$)} & $ |V| $ & $ |E| $  & Avg. degree & Diameter \\ \hline \hline
     \textbf{RoadNet}     &  56M    & 717M      & 1.05        & 48K  \\ \hline   	
   	 \textbf{DBLP}        &  0.3M   & 1.0M      & 6.62        & 21   \\ \hline
   	 \textbf{LiveJournal} &  4.8M   & 42.9M     & 18          & 17 \\ \hline
   	 \textbf{UK2002}      &  18.5M  & 298.1M    & 32          & 22 \\ \hline
	\end{tabular}
\end{table}

On disk, our data graphs are stored in plain text format where each line represents an adjacency-list of a vertex. The approach of \qiao relies on the clique-index of the data graph which should be pre-constructed and stored on disk. In Table~\ref{t:qiao_index}, we present the disk space cost of the index files generated by the program of \qiao (M for Mega Bytes, G for Giga Bytes).

\begin{table}[h]
    \caption{Illustration of the Size of Index Files of \qiao}
	\centering
	\small
	\label{t:qiao_index}
 	\begin{tabular}{|c|c|c|c|c|}
    \hline
     \textbf{Dataset($G$)} & Data Graph File Size & Index File Size \\ \hline \hline
   	 \textbf{DBLP}   & 13M   & 210M   \\ \hline
     \textbf{RoadNet} & 2.3G   & 16.9G     \\ \hline
   	 \textbf{LiveJournal}   & 501M   & 6.5G   \\ \hline
   	 \textbf{UK}   & 4.1G   & 60G   \\ \hline
	\end{tabular}
\end{table}

\begin{figure}[htbp!]
\small
\centering
\includegraphics[width=0.4\textwidth]{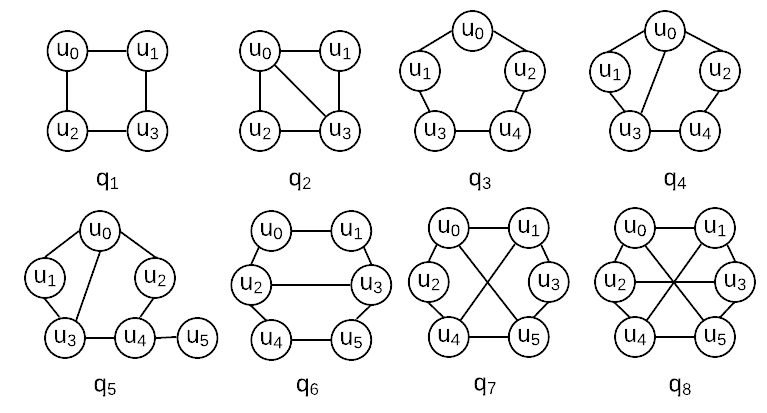}
\caption{Query Set}
\label{fig:query1}
\end{figure}

The queries we used are given in Figure~\ref{fig:query1}.

We evaluate the performance, measured by time elapsed and communication cost, of the five approaches in Section~\ref{subsec:performanceCom}. The cluster we used for this experiment consists of 10 nodes. Due to space limit, more experimental results, including execution plan evaluation and scalability test etc., are presented in Appendix~\ref{appendix:moreResults}.

\subsection{Performance Comparison}
\label{subsec:performanceCom}

\begin{figure}[htbp!]		
      \centering
      \begin{subfigure}[b]{0.45\textwidth}
    		\includegraphics[width=\textwidth]{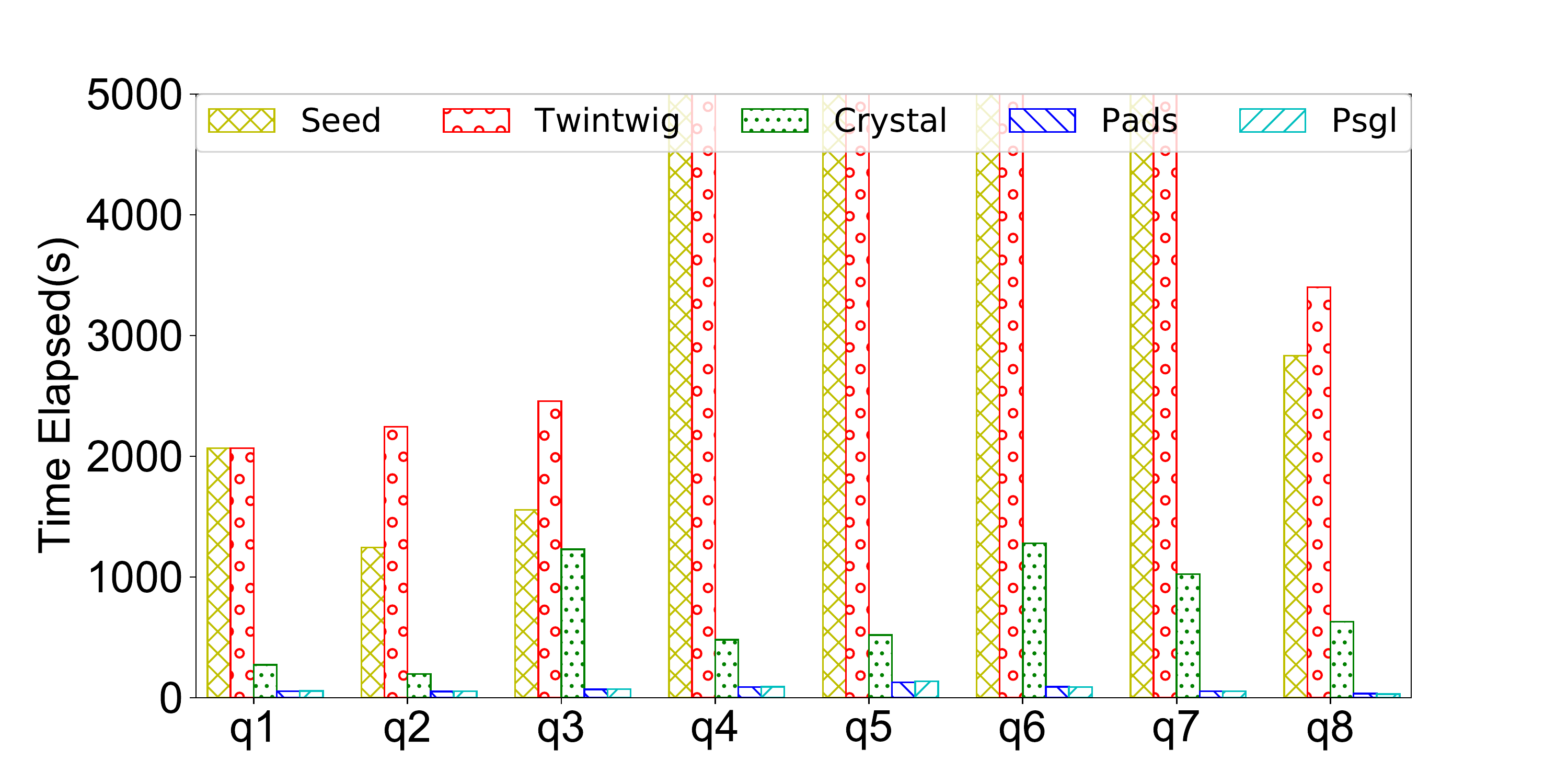}
		\caption{Time cost}
      \end{subfigure} \\%
        ~ 
      \begin{subfigure}[b]{0.45\textwidth}
    		\includegraphics[width=\textwidth]{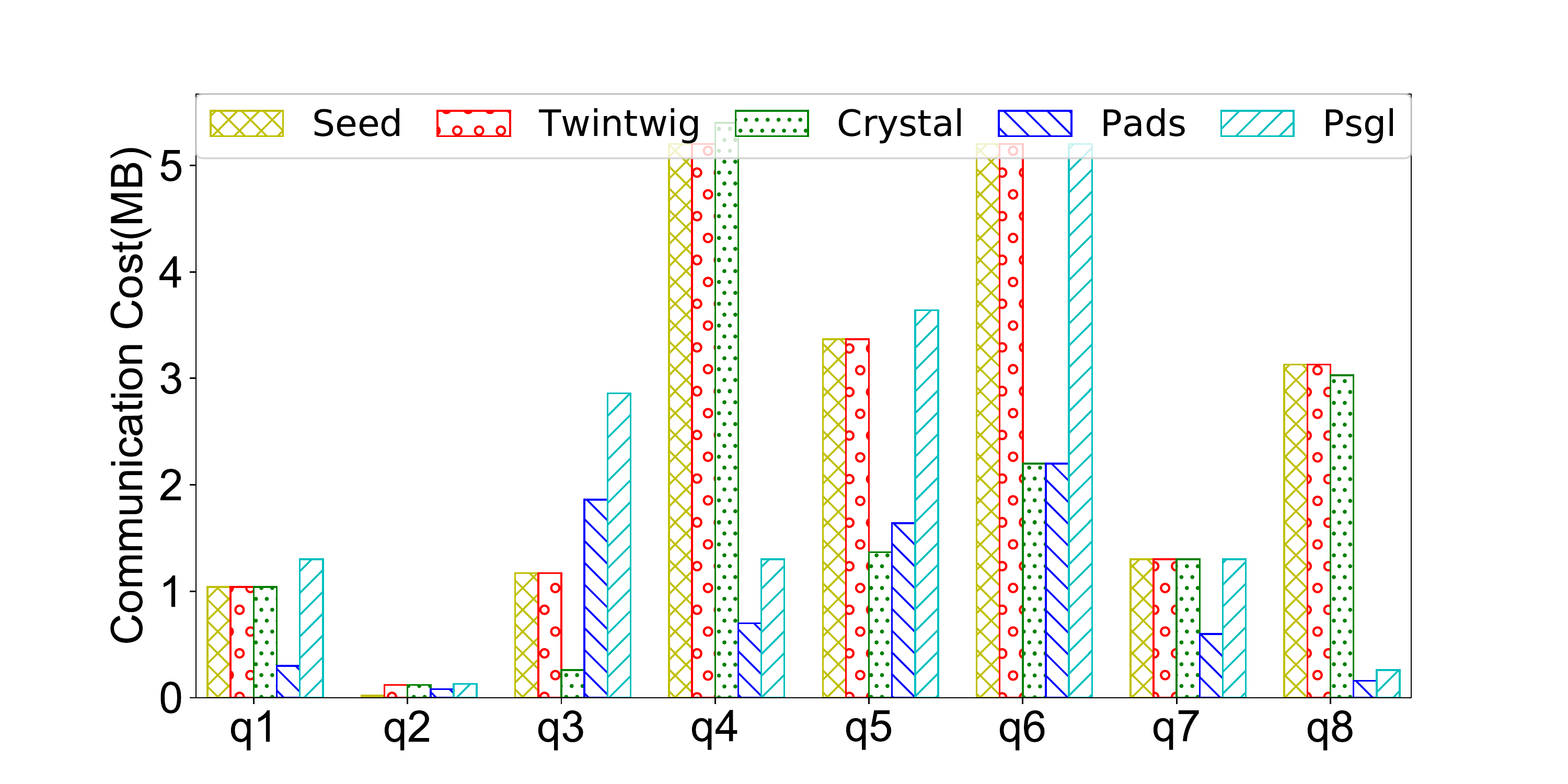}
         \caption{Communication Cost}
      \end{subfigure}
	\caption{Performance over RoadNet}
	\label{fig:10_roadnet}
\end{figure}

\begin{figure}[htbp!]		
      \centering
      \begin{subfigure}[b]{0.45\textwidth}
    		\includegraphics[width=\textwidth]{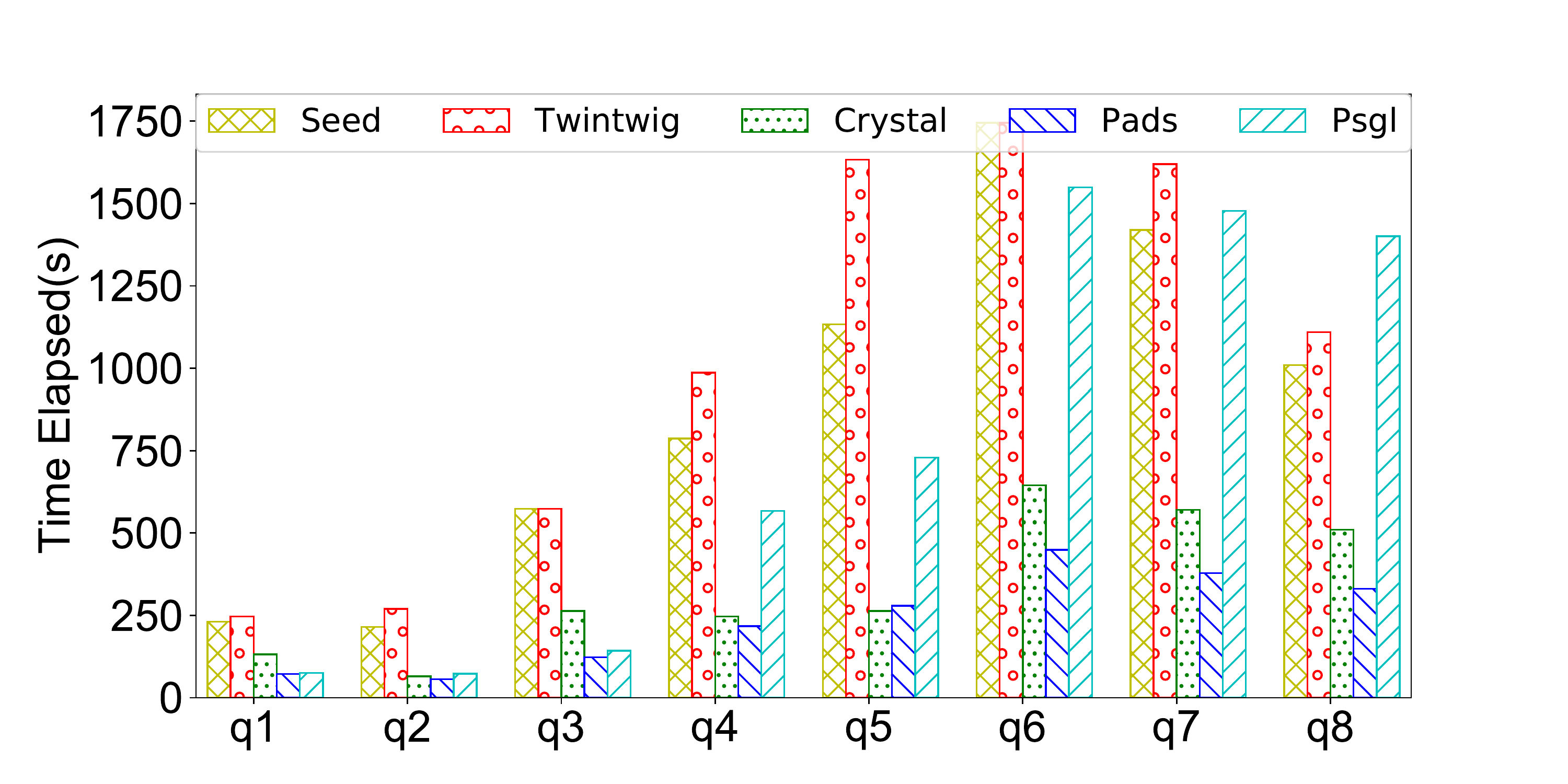}
		\caption{Time cost}
      \end{subfigure} \\%
        ~ 
      \begin{subfigure}[b]{0.45\textwidth}
    		\includegraphics[width=\textwidth]{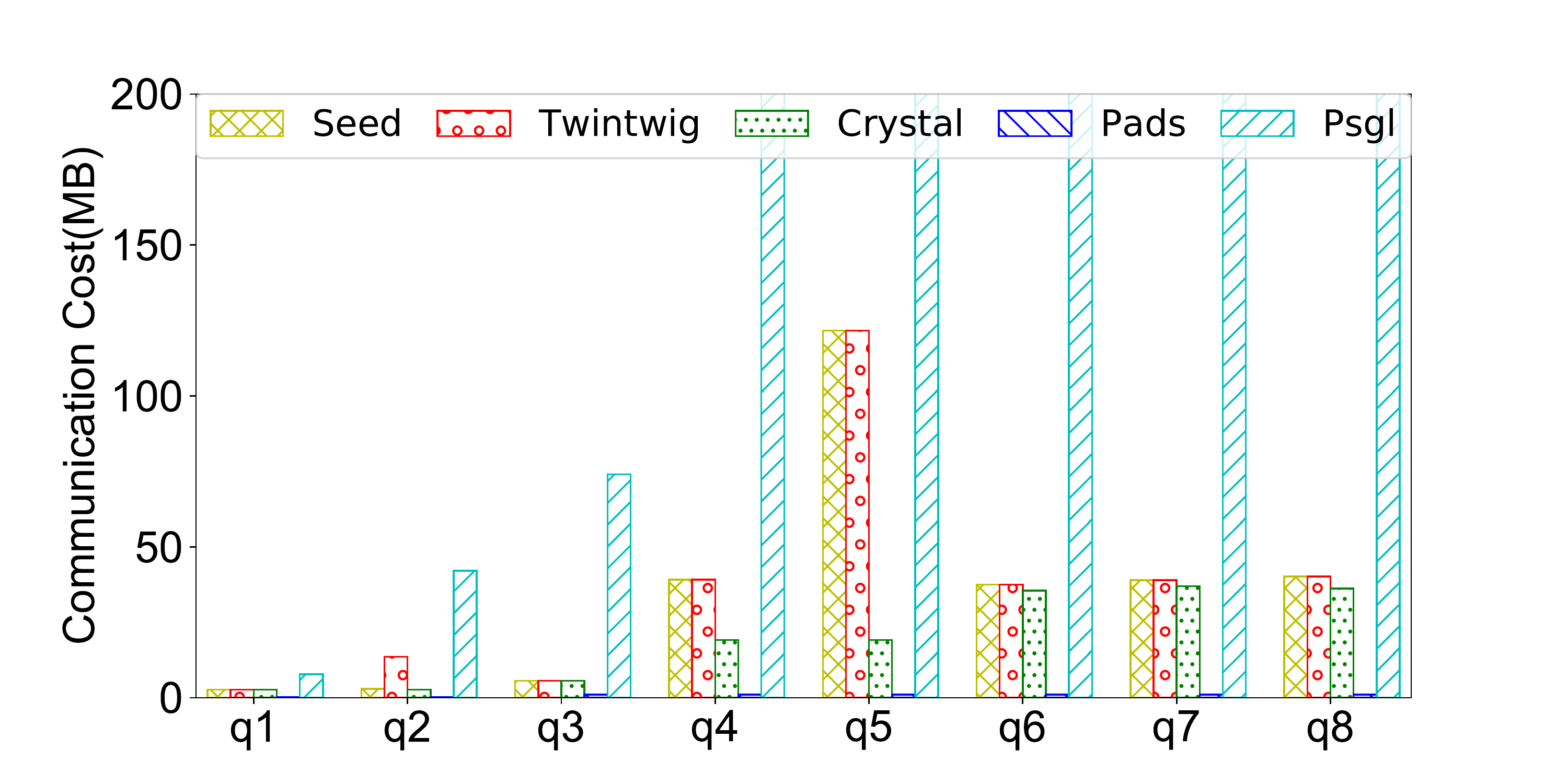}
         \caption{Communication Cost}
      \end{subfigure}

	\caption{Performance over DBLP}
	\label{fig:10_dblp}
\end{figure}

\begin{figure}[htbp!]		
      \centering
      \begin{subfigure}[b]{0.45\textwidth}
    		\includegraphics[width=\textwidth]{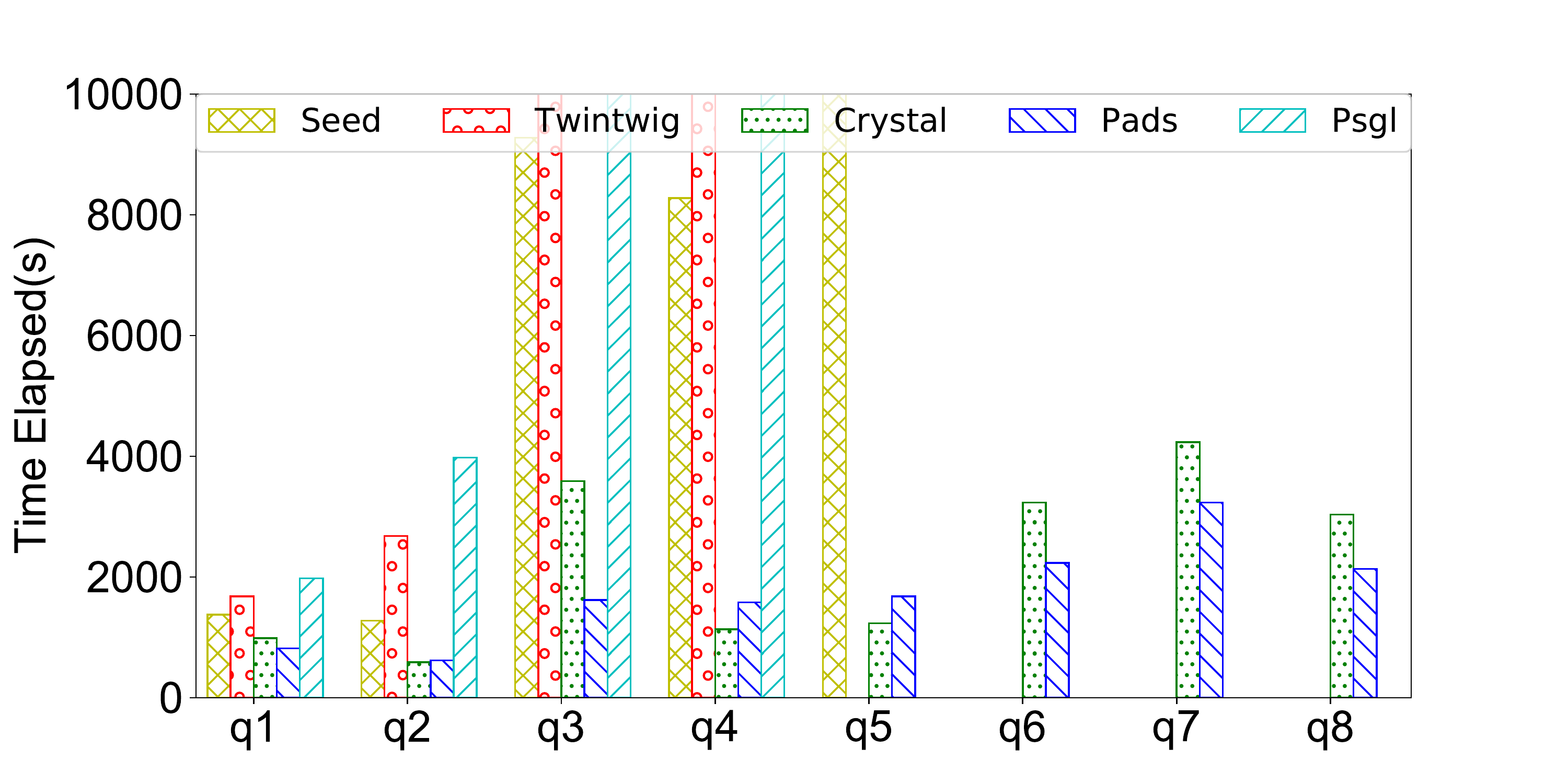}
		\caption{Time cost}
      \end{subfigure} \\%
        ~ 
      \begin{subfigure}[b]{0.45\textwidth}
    		\includegraphics[width=\textwidth]{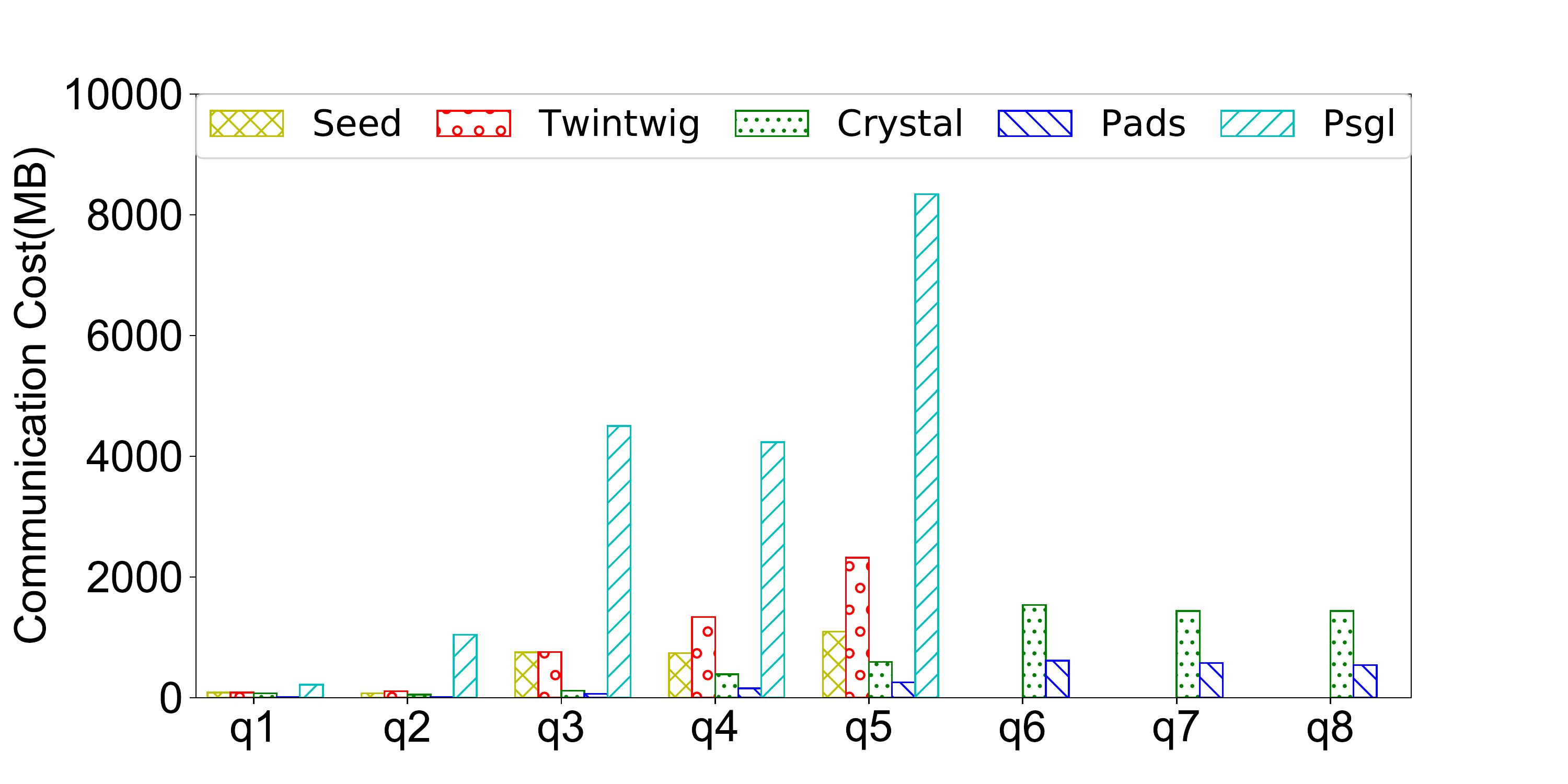}
         \caption{Communication Cost}
      \end{subfigure}

	\caption{Performance over LiveJournal}
	\label{fig:10_livejournal}
\end{figure}

\begin{figure}[htbp!]		
   \centering
   \includegraphics[width=0.45\textwidth]{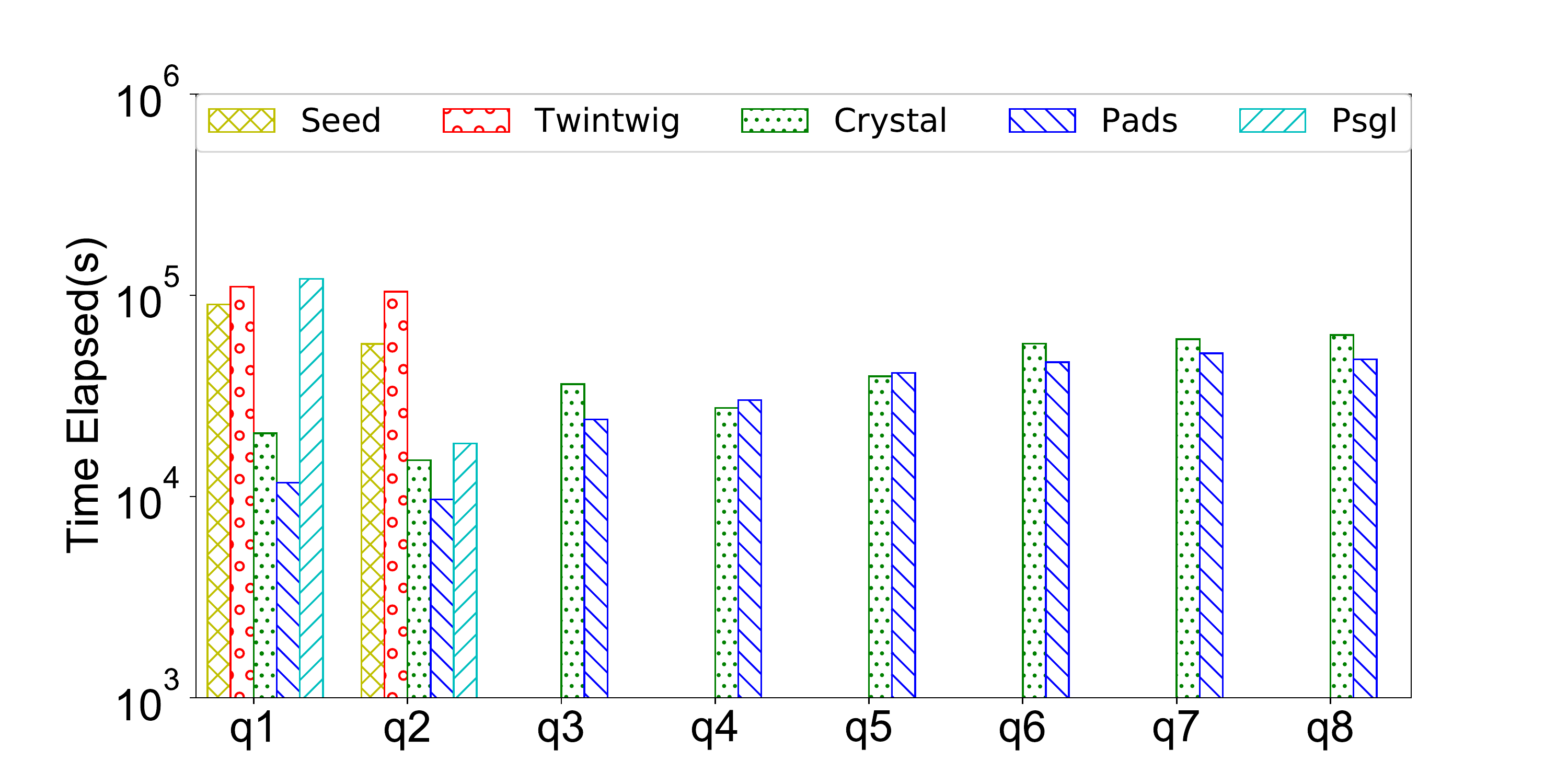}
   \caption{Time cost over UK2002}
   \label{fig:10_uk}
\end{figure}

We compare the performance of five subgraph enumeration approaches by measuring the time elapsed (in seconds) and the volume of exchanged data of processing each query pattern. The results of DBLP, RoadNet, LiveJournal and UK2002 are given in \ref{fig:10_roadnet}, Figure~\ref{fig:10_dblp},  Figure~\ref{fig:10_livejournal} and Figure~\ref{fig:10_uk} , respectively. We mark the result as empty when the test fails due to out-of-memory errors. When any bar reaches the upper bound, it means the corresponding values is beyond the upper bound value shown in the chart.

\stitle{Exp-1:RoadNet}
The results over the RoadNet dataset are given in Figure~\ref{fig:10_roadnet}. As can be seen from the figure, \Pads and \psgl are significantly faster than the other three methods (by more than 1 order of magnitude). \Pads and \psgl are using graph exploration while the others are using join-based methods. Therefore, both \Pads and \psgl demonstrated efficient filtering power. Since join-based methods need to group the intermediate results based on keys so as to join them together, the performance was significantly dragged down when dealing with sparse graphs compared with \Pads and \psgl.

It is worth noting that \psgl was verified slower than \twintwig and \seed in \cite{DBLP:journals/pvldb/LaiQLC15}\cite{DBLP:journals/pvldb/LaiQLZC16}. This  may be  because the datasets used in \twintwig and \seed are much denser than RoadNet, hence a huge number of embeddings will be generated. The grouped intermediate results of \twintwig and \seed significantly reduced the cost of network traffic. Another interesting observation is that although \qiao has heavy indexes, its performance is much worse than \psgl and \Pads.
The reason is that the number of cliques in RoadNet is relatively small considering the graph size.
\ignore{This indicates the performance of \qiao is highly affected by the type of data graphs.}
Moreover, there are no cliques with more than two vertices in queries $q_1$, $q_3$, $q_6$, $q_7$ and $q_8$. In such cases,
\ignore{the method of \qiao is similar to the other two join-methods with each decomposition unit being an edge.}
the clique index cannot help to improve the performance.

As shown in Figure~\ref{fig:10_roadnet}(b), the communication cost is not large for any of the approaches (less than 5M for most queries). In particular, for  \Pads, the communication cost is almost 0 which is mainly because most data vertices can be processed by \sme, as such no network communication is required.

\stitle{Exp-2:DBLP} The result over DBLP is shown in Figure~\ref{fig:10_dblp}.
As aforementioned, DBLP is smaller but much denser than RoadNet. The number of intermediate results generated in DBLP are much larger than that in RoadNet, as implied by the data communication cost shown in Figure~\ref{fig:10_dblp}~(b).
Since \psgl does not consider any compression or grouping over intermediate results, the communication cost of \psgl is much higher than the other approaches (more than 200M for queries after $q_4$).  Consequently, the time delay due to shuffling the intermediate results caused bad performance for \psgl. However, \psgl is still faster than \seed and \twintwig. This may be because the time cost of grouping intermediate results of \twintwig and \seed is heavy as well. It is worth noting that the communication cost of our \Pads is quite small (less than 5M). This is because of the caching strategy of \Pads where most foreign vertices are only fetched once and cached in the local machine. If most vertices are cached, there will be no further communication cost. The time efficiency of \Pads is better than \qiao even for queries $q_2$,$q_4$ and $q_5$ where the triangle crystal can be directly loaded from index without any computation.

\stitle{Exp-3:LiveJournal} As shown in Figure~\ref{fig:10_livejournal}, for LiveJournal, \seed, \twintwig and \psgl start becoming impractical for queries from $q_3$ to $q_8$. It took them more than 10 thousand seconds in order to process each of those queries. Due to the huge number of intermediate results generated, the communication cost increased significantly as well, especially for \psgl whose communication cost was beyond control when the query vertices reach 6. The method of \qiao achieved good performance for queries $q_2$, $q_4$ and $q_5$. This is mainly because \qiao simply retrieved the cached embeddings of the triangle to match the vertices $(u_0,u_1,u_2)$ of those 3 queries. However, when dealing with the queries with no good crystals ($q_6$, $q_7$ and $q_8$), our method significantly outperformed \qiao. One important thing to note is that the other three methods (\seed, \twintwig and \psgl) are sensitive to the end vertices, such as $u_5$ in $q_5$. Both time cost and communication cost increased significantly from $q_4$ to $q_5$. \Pads processes those end vertices last by simply enumerating the combinations without caching any results related to them. The end vertices within \qiao will be bud vertices which only requires simple combinations. As indicated by query $q_5$ where their processing time increased slightly from that of $q_4$, \Pads and \qiao are nicely tuned to handle end vertices.

\stitle{Exp-4:UK2002}
As shown in Figure~\ref{fig:10_uk}, \twintwig, \seed and \psgl failed the tests of queries after $q_3$ due to memory failure caused by huge number of intermediate results. The communication cost of all other methods are significantly larger than \Pads (more than 2 orders of magnitude), we omit the chart for communication cost here. Similar to that of LiveJournal, the processing time of \qiao is better than that of \Pads for the queries with cliques. This is because  \qiao directly retrieves the embeddings of the cliques from the index. However, for queries without good crystals, our approach demonstrates better performance. As shown in Table~\ref{t:qiao_index}, the index files of \qiao is more than 10 times larger than the original data graph. 

Another advantage of \Pads over \qiao is our memory control strategies ensures it is more robust: we tried to set a memory upper bound of 8G and test query $q_6$, \qiao starts crashing due to memory leaks,  while \Pads successfully finished the query for this test.

\section{Related Work}
\label{sec:relatedWork}

\ifpdf
    \graphicspath{{Chapter4/}}
\else
    \graphicspath{{Chapter4/}}
\fi

The works most closely related to ours are \twintwig \cite{DBLP:journals/pvldb/LaiQLC15}, \seed \cite{DBLP:journals/pvldb/LaiQLZC16} and \psgl \cite{DBLP:conf/sigmod/ShaoCCMYX14}. Both \cite{DBLP:journals/pvldb/LaiQLC15} and \cite{DBLP:journals/pvldb/LaiQLZC16} use multi-round two-way joins. \cite{DBLP:journals/pvldb/LaiQLC15} uses the same data partitioning as in our work, and it decomposes the query graph $P$ into a set of small trees ${dp_0, \ldots, dp_k}$ such that the union of these trees is equal to $P$. Since the decomposition units are trees, a set of embeddings of $dp_i$ can be obtained on each machine without consulting other machines, and the union of the embeddings on all machines is the set of all embeddings of $dp_i$ over $G$.
In the first round, the embeddings of $dp_0$ and $dp_1$ are joined to obtain that of $P_1$; in each subsequent round, the embeddings of $P_{i-1}$ and $dp_i$ are joined to obtain that of $P_i$. Since
%
%
the embeddings of $P_{i-1}$ on one machine must be joined with the embeddings of $dp_i$ on every machine,  all the intermediate results (i.e., embeddings of $P_{i-1}$ and $dp_i$) must be cached and then shuffled based on the {\em join key} and re-distributed to the machines. Synchronization is necessary since shuffling and re-distribution can only start when all machines have the intermediate results ready.  \cite{DBLP:journals/pvldb/LaiQLZC16} is similar to \cite{DBLP:journals/pvldb/LaiQLC15}, except that it allows decomposition units to be cliques as well as trees, and it uses bushy join rather than left-deep joins\footnote{There are independent optimization strategies in each paper, of course.}. To compute the intermediate results for these units, it adopts a slightly different data partition strategy: it uses star-clique-preserved partitions. Both \twintwig and \seed may generate huge intermediate results, and shuffling, re-distribution and synchronization cost a lost of time. Our approach is different in that we do not use joins, instead we use expand-verify-filter on each machine, as such we generate less intermediate results, and we do not need to re-distribute them to different machines.

\psgl \cite{DBLP:conf/sigmod/ShaoCCMYX14} is
based on Pregel \cite{DBLP:conf/podc/MalewiczABDHLC09}. It maps the query vertices one at a time following breath-first traversal, so that partial matches are expanded repeatedly until the final results are obtained. In this way it avoids explicit joins similar to our approach.
%
%
However, there are important differences between \psgl and our system (\Pads). (1)
In each step of expansion, \psgl needs to shuffle and send the partial matches (intermediate results) to other machines, while \Pads does not need to do so.  (2) \psgl stores each (partial) match as a node of a {\em static} result tree, while \Pads stores the results in a dynamic and compact data structure. (3) There is no memory control in \psgl.


Also closely related to our work are \cite{DBLP:conf/sigmod/FanXWYJZZCT17} and \cite{DBLP:conf/sigmod/FanLLXYYX18}, which introduce systems for parallelizing serial graph algorithms, including (but not limited to) subgraph isomorphism search algorithms. These systems partition the data graph into different machines, but do not partition the query graph.
Each machine evaluates the query pattern on its own machine using a serial algorithm (e.g., VF2) independently of others, but before that it must copy parts of the data graph from other machines. These parts of the graph are determined as follows. For each  {\em boundary} vertex $v$ on the current machine, it copies the nodes and edges within a distance $d$ from $v$, where $d$ is the diameter of the query graph. The final results are obtained by collecting the final results from all machines.
Obviously, if the query graph diameter is large, and the data graph diameter is small (e.g., those of social network graphs), or there are many boundary vertices involved,
then the entire partition of the neighboring machine may have to be fetched. This will generate heavy network traffic as well as burden on the memory of the local machine.

The work \cite{DBLP:conf/icde/AfratiFU13} treats the query pattern as a conjunctive query, where each predicate represents an edge, and computes the results as a multi-way join in a single round of {\em map} and {\em reduce}. As observed in \cite{DBLP:journals/vldb/LaiQLC17}, the problem with this approach is that most edges have to duplicated over several machines in the map phase, hence there is a scalability problem when the query pattern is complex.

Qiao {\em et al} \cite{DBLP:journals/pvldb/QiaoZC17} represent the set $I_P$ of all embeddings of pattern $P$ in a compressed form, $code(I_P)$,  based on a minimum vertex cover of $P$. It decomposes the query graph $P$ into a core $core(P)$ and a set of so-called {\em crystals} $\{p_1, \ldots, p_k\}$, such that $code(I_P)$ can be obtained by joining the compressed results of $core(P)$ and $\{p_1, \ldots, p_k\}$. This join process can be parallelized in map-reduce. The compressed results of $core(P)$ and the crystals can be obtained from the compressed results of components of $P$.  To expedite query processing, it builds an index of all cliques of the data graph, {  as shown in Table~\ref{t:qiao_index}}. Although no shuffling of intermediate results is required, the indexes of \cite{DBLP:journals/pvldb/QiaoZC17} can be many times larger than the data graph, and computing/maintaining such big indexes can be very expensive, making it less practical. 

BigJoin, one of the algorithms proposed in \cite{DBLP:journals/pvldb/AmmarMSJ18}, treats a subgraph query as a  join of $|E_P|$ binary relations where each relation represents an edge in $P$. Similar to \Pads and \psgl, it generates results by expanding partial results a vertex at a time, assuming a fixed order of the query vertices. BigJoin targets achieving {\em worst-case} optimality. \ignore{It also controls memory usage by expanding the partial results a batch at a time, without detailing how the batches are chosen.} Different from our work, it still needs to shuffle and exchange intermediate results, and therefore synchronization before that.

\section{Conclusion}
\label{sec:conclusion}
We presented a practical asynchronous subgraph enumeration system $\Pads$ whose core is based on a new framework \textbf{R-Meef}(\rmeef). By processing the data vertices far away from the border using the single-machine algorithms, we isolated a large part of vertices which does not have to involve in the distributed process.
By passing verification results of foreign edges and adjacency-list of foreign vertices, \Pads significantly reduced the network communication cost. 
We also proposed a compact format to store the generated intermediate results.
Our query execution plan and several memory control strategies including foreign vertex caching and region groups are designed to improve the efficiency and robustness of \Pads. 
Our experiment results have verified the superiority of \Pads compared with state-of-the-art subgraph enumeration approaches.

\bibliographystyle{abbrv}
\bibliography{Reference/reference}  

\appendix

\section{Proofs}
\subsection{Proof of Proposition \ref{lem:diaEmbed}}
\label{subsec:spanProof}

\begin{proof}
Suppose there is an embedding $f$ such that $f(u)=v$, $f(u')=v'$. We show $dist(v,v') \leq \BD_{G_t}(v)$, therefore $v'$ must be on $M_t$. Any shortest path from $u$ to $u'$ will be mapped by $f$ to a path in $G$, therefore $dist(v,v')\leq dist(u,u')$ $\leq\QD_{P}(u)$. By assumption, $\QD_{P}(u)\leq \BD_{G_t}(v)$, therefore  $dist(v,v')\leq \BD_{G_t}(v)$.
\end{proof}

\subsection{Proof of Theorem \ref{the:miniPlan}}
\label{subsec:planProof}

\begin{proof}
Suppose $\{dp_0, \ldots, dp_k\}$ is an execution plan. The plan has $k+1$ decomposition units. \ignore{According to Definition~\ref{defn:executionPlan},} Clearly the pivot vertices of the decomposition units form a connected dominating set of $\sPattern$. Therefore, $k+1 \geq c_{\sPattern}$. This proves any execution plan has at least $c_{\sPattern}$ decomposition units.

Now suppose $T$ is a \mlst  of $\sPattern$. From $|V_{\sPattern}|=c_{\sPattern}+l_{\sPattern}$ we know the number of non-leaf vertices in $T$ is $c_{\sPattern}$. We can construct an execution plan by choosing one of the non-leaf vertices $v_0$ as $dp_0.\piv$,
and all neighbors of $v_0$ in $T$ as the vertices in $dp_0.\lf$. Regarding $v_0$ as the root of the spanning tree $T$, we then choose each of the non-leaf children $v_i$ of $v_0$ in $T$ as the pivot vertex of the next decomposition unit $dp_i.\piv$,
and all children of $v_i$ as the vertices in $dp_i.\lf$. Repeat this process until every non-leaf vertex of $T$ becomes the pivot vertex of a decomposition unit. This decomposition has exactly $c_P$ units, and it forms an execution plan. This shows that there exists an execution plan with $c_{\sPattern}$ decomposition units.
%
\end{proof}

\section{Implementation of \Mur}
\label{sec:rmeefImplementaion}
We present the implementation of $\Mur$ as shown in Algorithm~\ref{algorithm:disSubEnum}.
\begin{algorithm}
\DontPrintSemicolon
\SetAlgoLined
\KwIn{Query pattern $P$,  partition $G_t$ on machine $M_{t}$, execution plan $\joinplan$}
\KwOut{$\mathbb{R}_{G_t}(P)$}

	$\regiongroup$ $=$ $\{rg_0 \dots rg_k\}$ $\gets$    $regionGroups\big(C(dp_0.\piv, M_t)\big)$ \\
		
	\For{\textbf{each} region group $rg \in $ $\regiongroup$}
	{	
		init \et $\eTrie$ with size $|V_{P}|$\\
	
		init edge verification index $\ei$ \\
	
		\For{\textbf{each} data $v \in rg$}
		{
		    $f \gets (dp_0.\piv, v)$ \\
		    $\ei$ $\gets$ $expandEmbedTrie(f, M_t, dp_0, \eTrie)$ \\
		}
	
		$\mathcal{R} \gets verifyForeignE(\ei)$ \\
		$filterFailedEmbed(\mathcal{R}, \ei, \eTrie)$ \\
	
		\For{Round $i = 1$ to $|\joinplan|$}
		{
	    		 clear $\ei$ \\

			$fetchForeignV$($i$) \\
	
			\For{\textbf{each} $f \in $ $\ei$}
			{			
					$\ei$ $\gets$ $expandEmbedTrie(f, M_t, dp_i, \eTrie)$ \\
			}
		
			$\mathcal{R} \gets verifyForeignE(\ei)$      \\
			$filterFailedEmbed(\mathcal{R}, \ei, \eTrie)$   \\
		}
		$\mathbb{R}_{G}(P)$ $\gets$ $\mathbb{R}_{G}(P)$ $\cup$ $\eTrie$ \\
		clear $\eTrie$ \\
	}
\BlankLine
\caption{\sc \Mur Framework}
\label{algorithm:disSubEnum}
\end{algorithm}

\ignore{Within each machine, we first compute an execution plan $\joinplan$ based on the given stripped query pattern $\sPattern$ (Line 1, 2). The execution plans computed in every machine are the same.}
Within each machine, we group the candidate data vertices of $dp_0.\piv$ within $M_t$ into  region groups (Line 1).
For each region group $rg$, a multi-round mapping process is conducted (Line 2 to 18).
Within each round, we use a data structure $\eTrie$ (embedding trie) to save the generated intermediate results, i.e., embeddings and embedding candidates (Line 3).
The edge verification index $\ei$ is initialized in Line 4, which will be reset for each round of processing (line 11).

\begin{enumerate}[leftmargin=*]
\item[(1)] {\textit{First Round (round 0)}}  Starting from each candidate $v$ of  $rg$, we match $v$ to $dp_0.\piv$ in the execution plan. After the pivot vertex is matched, we find all the \ec{s} of $dp_0$ with respect to $M_t$ and compress them into $\eTrie$.  We use a function $expandEmbedTrie$ to represent this process (Line 7).  For each \ec compressed in $\eTrie$, its undetermined edges need to be verified in order to determine whether this \ec is an embedding of $dp_0$. We record this information in the edge verification index $\ei$, which is constructed in the $expandEmbedTrie$ function.  After we have the EVI $\ei$ in $M_t$, we send a $verifyE$ request to verify those undetermined edges within $\ei$ in the machine which has the ability to verify it (function $verifyForeignE$ in Line 8).  After the edges in $\ei$ are all verified, we remove the failed \ec{s} from $\eTrie$ (Line 9).

\item[(2)] {\textit{Other Rounds}}
For each of the  remaining rounds of the execution plan, we first clear the EVI  $\ei$ from previous round (Line 11). In the $i^{th}$ round, we want to find all the \ec{s} of $P_i$ based on the embeddings in $R_{G_t}(P_{i-1})$ (where  $dp_i.\piv$ has been matched). The process is to expand every embedding $f$ of $R_{G_t}(P_{i-1})$ with each embedding { candidate} of $dp_i$
within the neighbourhood of $f(dp_i.\piv)$.
 If not all the data vertices  matched to $dp_i.\piv$ by the {\ec}s in $R_{G_t}(P_{i-1})$ reside in $M_t$, we will have to fetch the adjacency-lists of those foreign vertices from other machines in order to expand from them. A sub-procedure $fetchForeignV$ is used to represent this process (Line 12).  After fetching, for each embedding $f$ of $R_{G_t}(P_{i-1})$, we find all the \ec{s} of $P_i$ by expanding from $f(dp_i.\piv)$ (Line 14). The found \ec{s} are compressed into $\eTrie$. Then $veriForeignE$ and $filterFailedEmbed$ are called to make sure that the failed \ec{s} are filtered out from the embedding trie,  which will only contain the actual embeddings of $P_i$, i.e.,  $R_{G_t}(P_{i})$ (Line 15, 16).

\end{enumerate}

After all the rounds of this region group have finished, we have a set of embeddings of $\sPattern$ compressed into $\eTrie$. The results obtained from all the region groups are put together to obtain the embeddings found by $M_t$. 
%

One important thing to note is that if a foreign vertex is already cached in the local machine, for the undetermined edges attached to this vertex, we can verify them locally without sending requests to other machines. Also we do not re-fetch any foreign vertex if it is already cached previously.

\begin{exmp}
\label{exmp:runningExmp}
Consider the data graph $G$ in Figure~\ref{fig:runningExample1}, where the vertices marked with dashed border lines reside in $M_1$ and the other vertices reside in $M_2$. 
Consider the  pattern $P$ and execution plan $\joinplan$ given in Example~\ref{exmp:executionPlan}.
We assume the preserved orders due to symmetry breaking are: $u_1<u_2$, $u_3$ $<$ $u_6$, $u_4$ $<$ $u_5$ and $u_8 < u_9$.

There are two vertices $\{v_0$, $v_{2}\}$ in $M_0$ and two vertices $\{v_1$, $v_{10}\}$ in $M_2$ with a degree not smaller than that of $dp_0.\piv$. Therefore in $M_1$, we have $C(dp_0.\piv)$ $=$ $\{v_0$, $v_2\}$ and in $M_2$ we have $C(dp_0$.$\piv)$ $=$ $\{v_1$, $v_{10}\}$.  After grouping, assume we have $\regiongroup=\{rg_0$, $rg_1\}$ where $rg_0$ $=$ $\{v_0\}$ and $rg_1$ $=$ $\{v_2\}$ in $M_1$,  and  $\regiongroup=\{rg_0\}$ where $rg_0=\{v_1$, $v_{10}\}$ in $M_2$.

Consider the region group $rg_0$ in $M_1$. In round $0$, we first match $v_0$ to $dp_0.\piv$. Expanding from $v_0$, we may have \ec{s} including by not limit to (we lock $u_7$ to $v_7$ for easy demonstration):

$f_{G_1}$ $=$ $\{(u_0$, $v_0)$, $(u_1$, $v_1)$, $(u_2$, $v_2)$, $(u_7$, $v_7)\}$

$f_{G_1}'$ $=$ $\{(u_0, v_0)$, $(u_1$, $v_1)$, $(u_2$, $v_{9})$, $(u_7$, $v_7)\}$

$f_{G_1}''$ $=$ $\{(u_0, v_0)$, $(u_1$, $v_9)$, $(u_2$, $v_{11})$, $(u_7$, $v_7)\}$

\noindent We compress these \ec{s} into $\eTrie$. Note that a mapping such as $\{(u_0, v_0)$, $(u_1$, $v_1)$, $(u_2$, $v_{11})$, $(u_7$, $v_7)\}$ is not an \ec of $dp_0$ $w.r.t$ $M_1$ since $(v_1$, $v_{11})$ can be locally verified to be non-existent.
Since the undermined edge $(v_1$, $v_{9})$ of $f_{G_1}'$ cannot be determined in $M_1$, we put $\{(v_1, v_{9}), <f_{G_1}'>\}$ into the EVI \ei.
We then ask $M_2$ to verify the existence of the edge. $M_2$ returns false, therefore $f_{G_1}'$ will be removed from $\eTrie$.

In round $1$, we have two embeddings $R_{G_t}(P_0)$= $\{f_{G_1}$, $f_{G_1}''\}$ to start with. To extend $f_{G_1}$ and $f_{G_1}''$, we need to fetch the adjacency-lists of $v_1$ and $v_{9}$ respectively. We send a single $fetchV$ request to fetch the adjacency-lists of $v_1$ and $v_{9}$ from $M_2$. After expansion from $v_1$, we get a single embedding $\{(u_0$, $v_0)$, $(u_1$, $v_1)$, $(u_2$, $v_2)$, $(u_3$, $v_3)$, $(u_4$, $v_4)$, $(u_7$, $v_7)$ $\}$ in $R_{G_t}(P_1)$. There is no embedding of $P_1$ expanded from $v_{9}$. Hence $f_{G_1}''$ will be removed from the embedding trie.


In round 2,  we expand from $v_2$ to get the \ec{s} of $P_2$. {$dp_2.piv$ was already mapped to $v_2$ as seen above, and $v_2$ has neighbors $v_5, v_6$ and $v_{10}$ that are not matched to any query vertices. Since there are sibling edge $(u_5,u_6)$ and cross-unit edge $(u_4, u_5)$ in $P_2$, we need to verify the existence of $(v_4, v_5)$ and $(v_5, v_6)$ if we want to map $u_5$ to $v_5$ and map $u_6$ to $v_6$. The existence of both $(v_4$, $v_5)$ and $(v_5, v_6)$ can be verified locally.
Similarly if we want to map $u_5$ to $v_{5}$, $u_6$ to $v_{10}$, we will have to verify the existence of $(v_5, v_{10})$, and so on. }
{It can be locally verified that $(v_5, v_{10})$ does not exist, and remotely verified that $(v_6, v_{10})$ does not exist.} Therefore, at the end of this round, we will get a single embedding for $P_2$ { which extends the embedding for $P_1$ by mapping $u_5, u_6$ to $v_5, v_6$ respectively.} We expand the embedding trie accordingly.

Following the above process, after we process the last round, we have an embedding of $P$ starting from region group $rg_0$ in machine $M_1$ will be saved in $\eTrie$:

$f_{G_1}$ $=$ $\{(u_0, v_0)$, $(u_1$, $v_1)$, $(u_2$, $v_{2})$, $(u_3$, $v_{3})$, $(u_4$, $v_{4})$, $(u_5$, $v_{5})$, $(u_6$, $v_{6})$, $(u_7$, $v_{7})$, $(u_8$, $v_{9})$, $(u_9$, $v_{11})\}$

\label{example3}
\end{exmp}

\section{More Experimental Results}
\label{appendix:moreResults}

\subsection{Scalability Test}
\label{appendix:scalaTest}
\begin{figure}[htbp!]		
      \centering
      \begin{subfigure}[b]{0.25\textwidth}
   		\includegraphics[width=\textwidth]{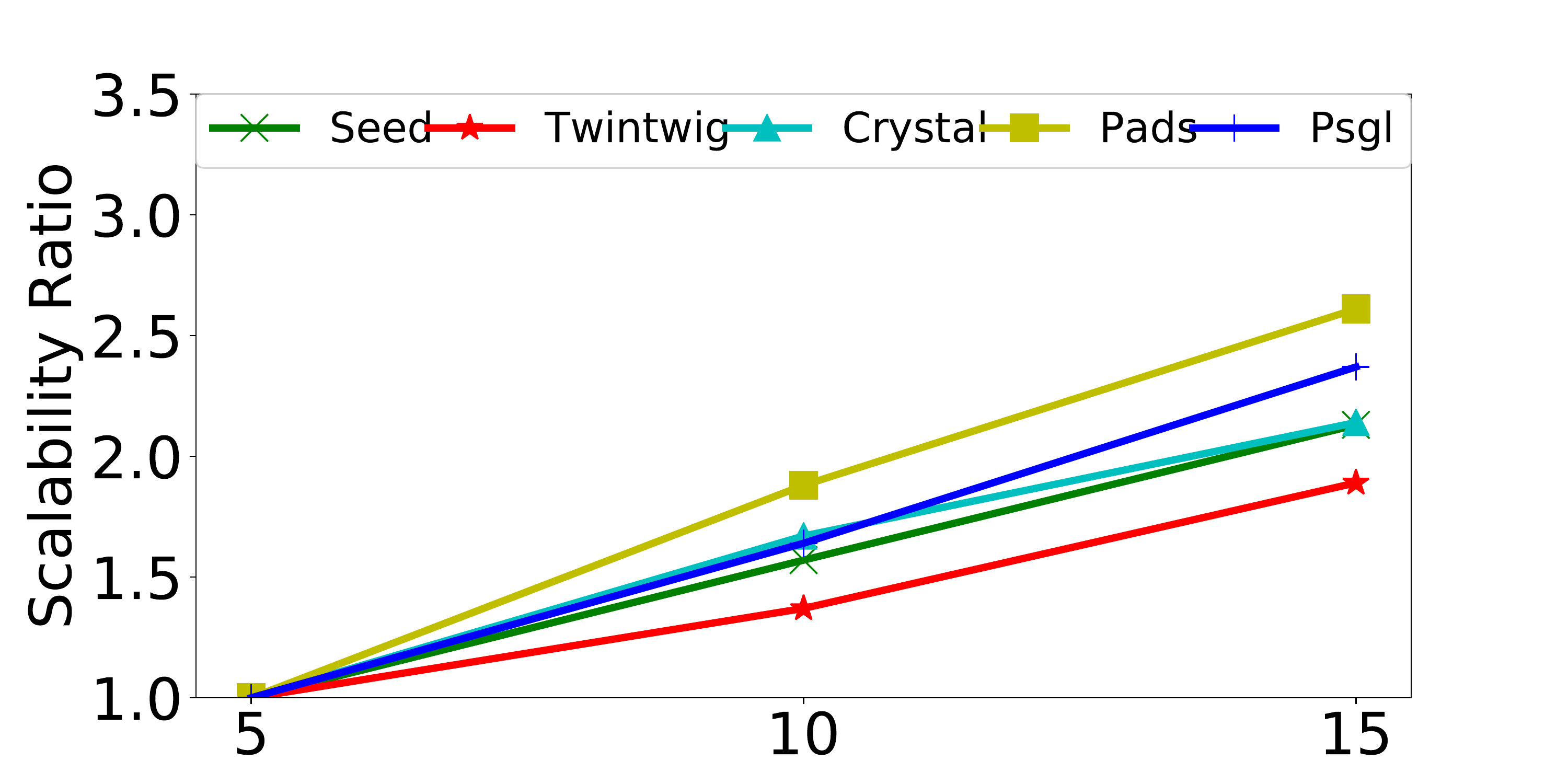}
		\caption{Roadnet}
      \end{subfigure}%
        ~ 
      \begin{subfigure}[b]{0.25\textwidth}
         \includegraphics[width=\textwidth]{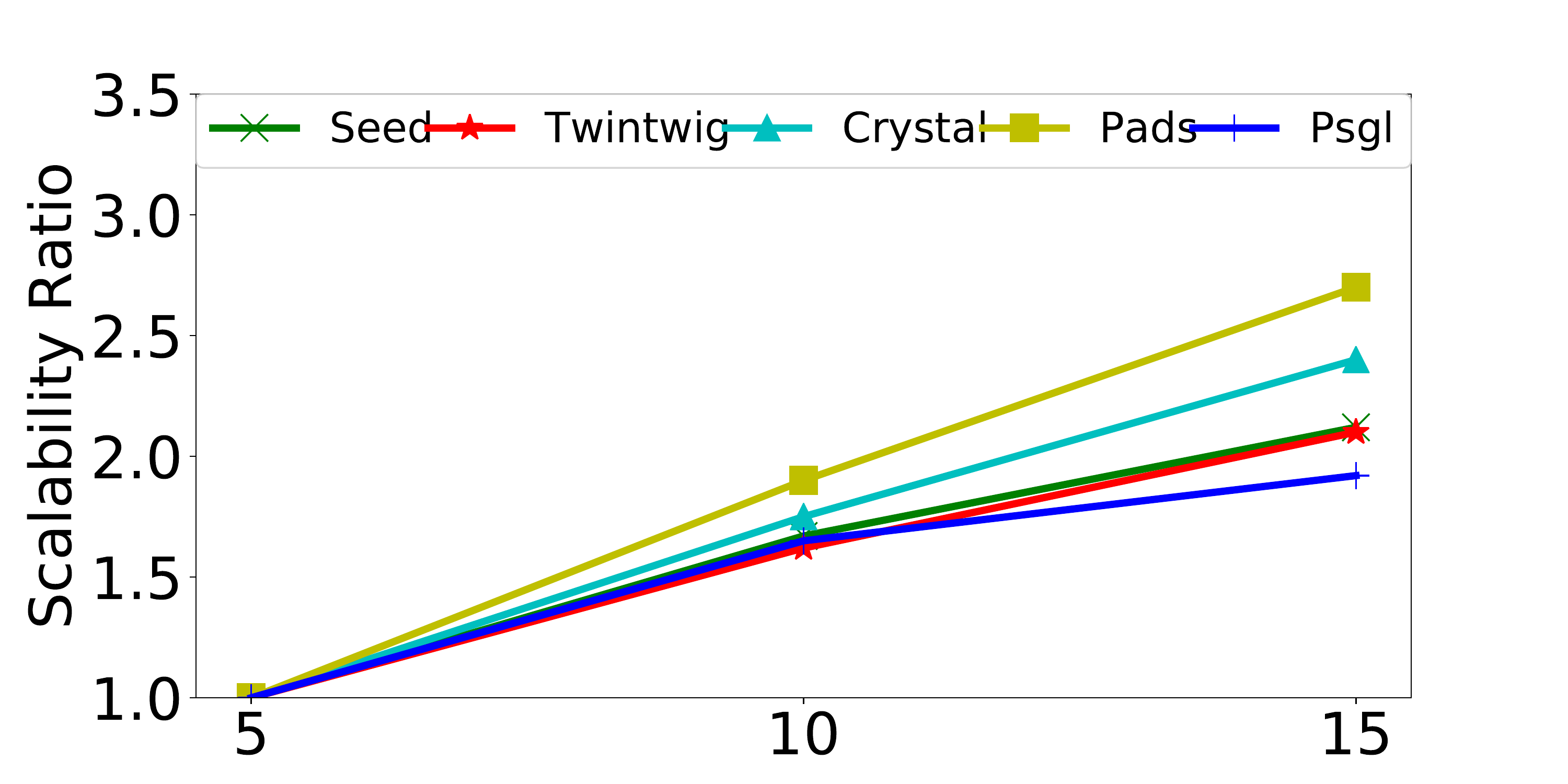}
         \caption{DBLP}
      \end{subfigure}

      \begin{subfigure}[b]{0.25\textwidth}
   		\includegraphics[width=\textwidth]{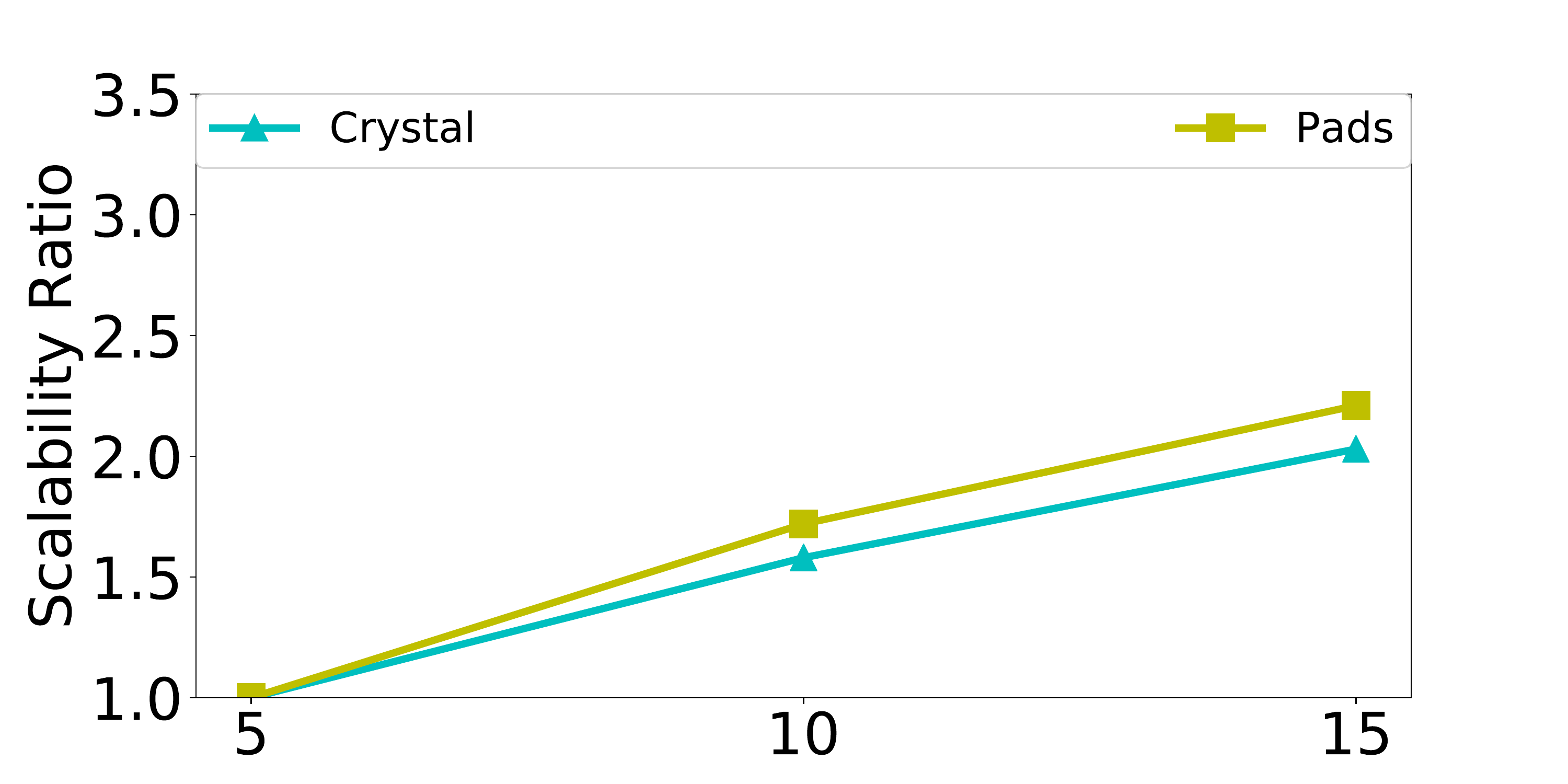}
		\caption{LiveJournal}
      \end{subfigure}%
        ~ 
      \begin{subfigure}[b]{0.25\textwidth}
         \includegraphics[width=\textwidth]{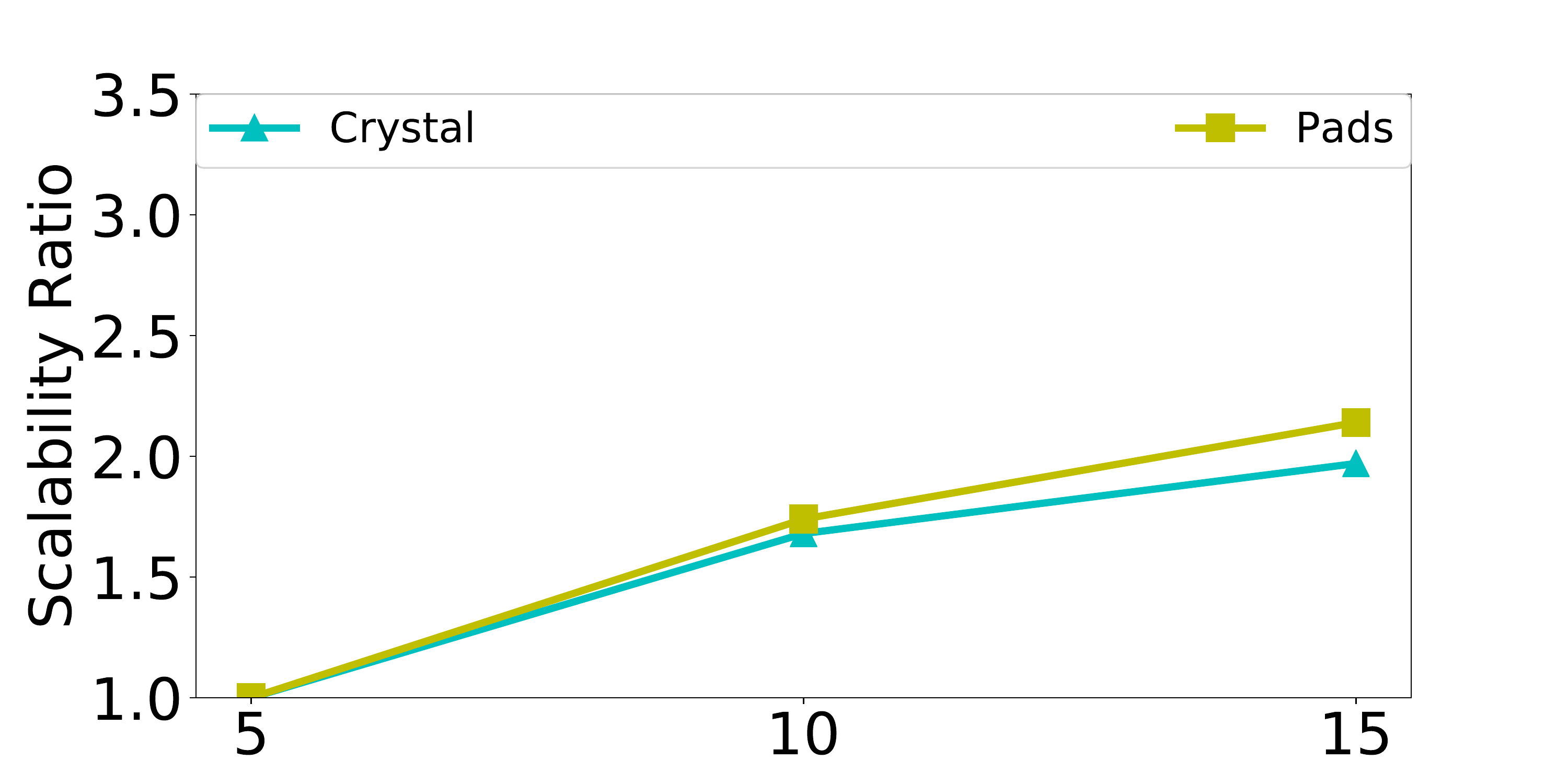}
         \caption{UK2002}
      \end{subfigure}

	\caption{Scalability Test}
	\label{fig:scalaTest}
\end{figure}

We compare the scalability of the five approaches by varying the number of nodes in the cluster (5, 10, 15), 3 cases in total. The queries we processed are shown in Figure~\ref{fig:query1}.
Instead of reporting the processing time, here we report the ratio between the total processing time of all queries using 5 nodes and that of the other two cases, which we call {\em scalability ratio}. \ignore{The gradient of each approach will indicate its scalability  in terms of utilizing cluster computing resources.} The results are as shown in Figure~\ref{fig:scalaTest}.

The most important thing to observe is that our approach demonstrates linear speed-up when the number of nodes is increased for Roadnet and DBLP.  The reason for Roadnet is because most vertices of each partition are far away from the border, therefore the majority of embeddings can be found by \sme. Each machine of our approach are almost independent except for some workload sharing. As for DBLP, which is a small graph, almost all vertices can be cached in memory, \Pads takes full advantage of it. \ignore{where it looks like each machine handles the same portion of data candidates.} Because \twintwig, \seed and \psgl failed some queries for LiveJournal and UK2002, we omit their scalability results in those two datasets.  The difference between \qiao and \Pads is not much while \Pads is better for both.

\subsection{Effectiveness of Query Execution Plan}
To validate the effectiveness of our strategy for choosing query execution plan, we compare the processing time of \Pads with two other baseline plans which are generated by replacing the execution plan of \Pads with the execution plans $RanS$ and $RanM$, respectively.
$RanS$ represents a plan consisting of random star decomposition units (no limit on the size of the star) and $RanM$ represents plan with minimum number of rounds without  considering the strategies in Section~\ref{subsec:movingFV}.
The cluster we used for this test consists of 10 nodes. In order to cover more random query plans, we run each test 5 times and report the average.
The queries are as shown in Figure~\ref{fig:query1}. For queries $q_1$ to $q_3$, the query plans generated in the above three implementations are almost the same. Therefore, we omit the data for those three queries.

\begin{figure}[htbp!]		
      \centering
      \begin{subfigure}[b]{0.25\textwidth}
   		\includegraphics[width=\textwidth]{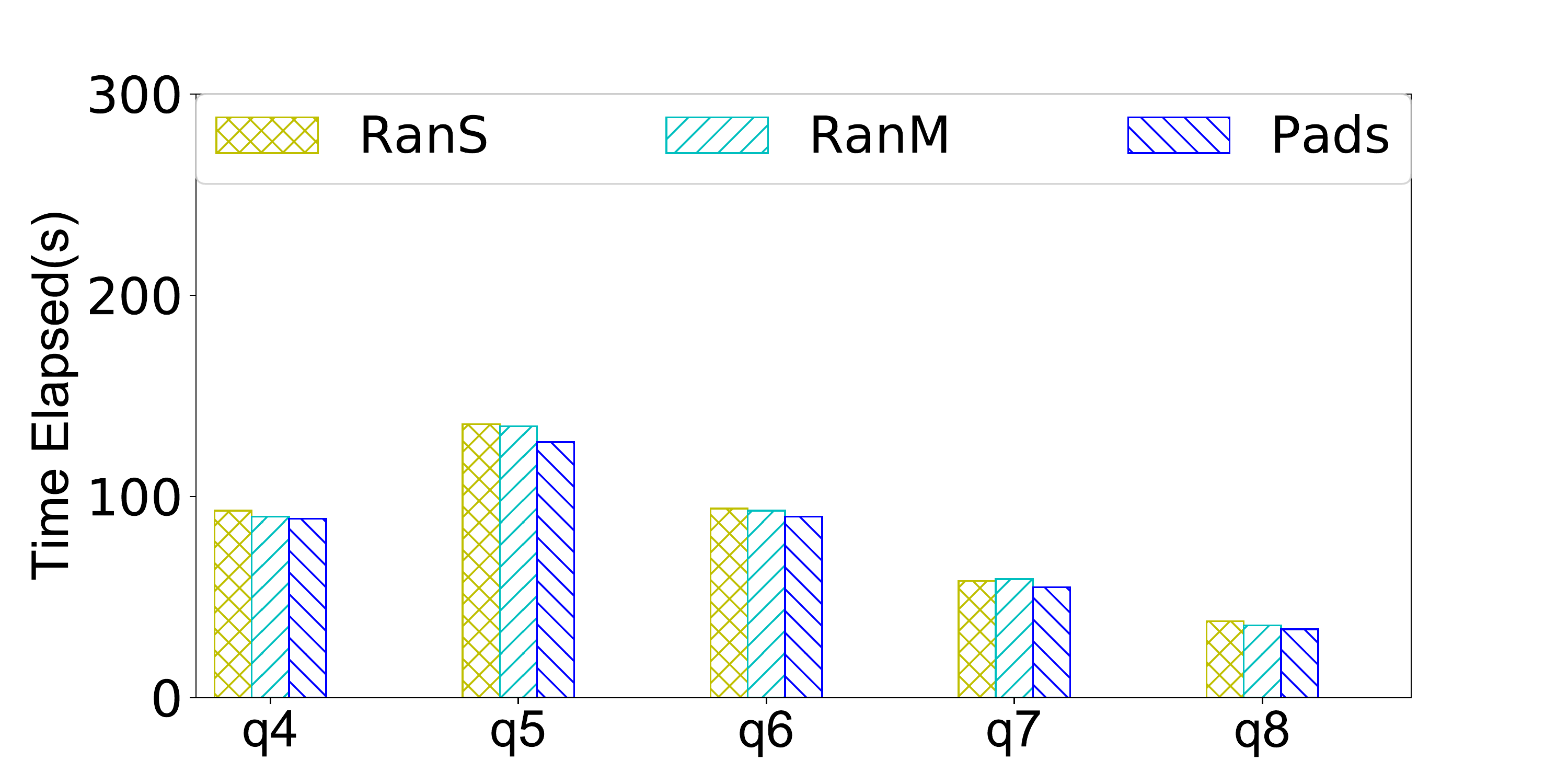}
		\caption{Roadnet}
      \end{subfigure}%
        ~ 
      \begin{subfigure}[b]{0.25\textwidth}
         \includegraphics[width=\textwidth]{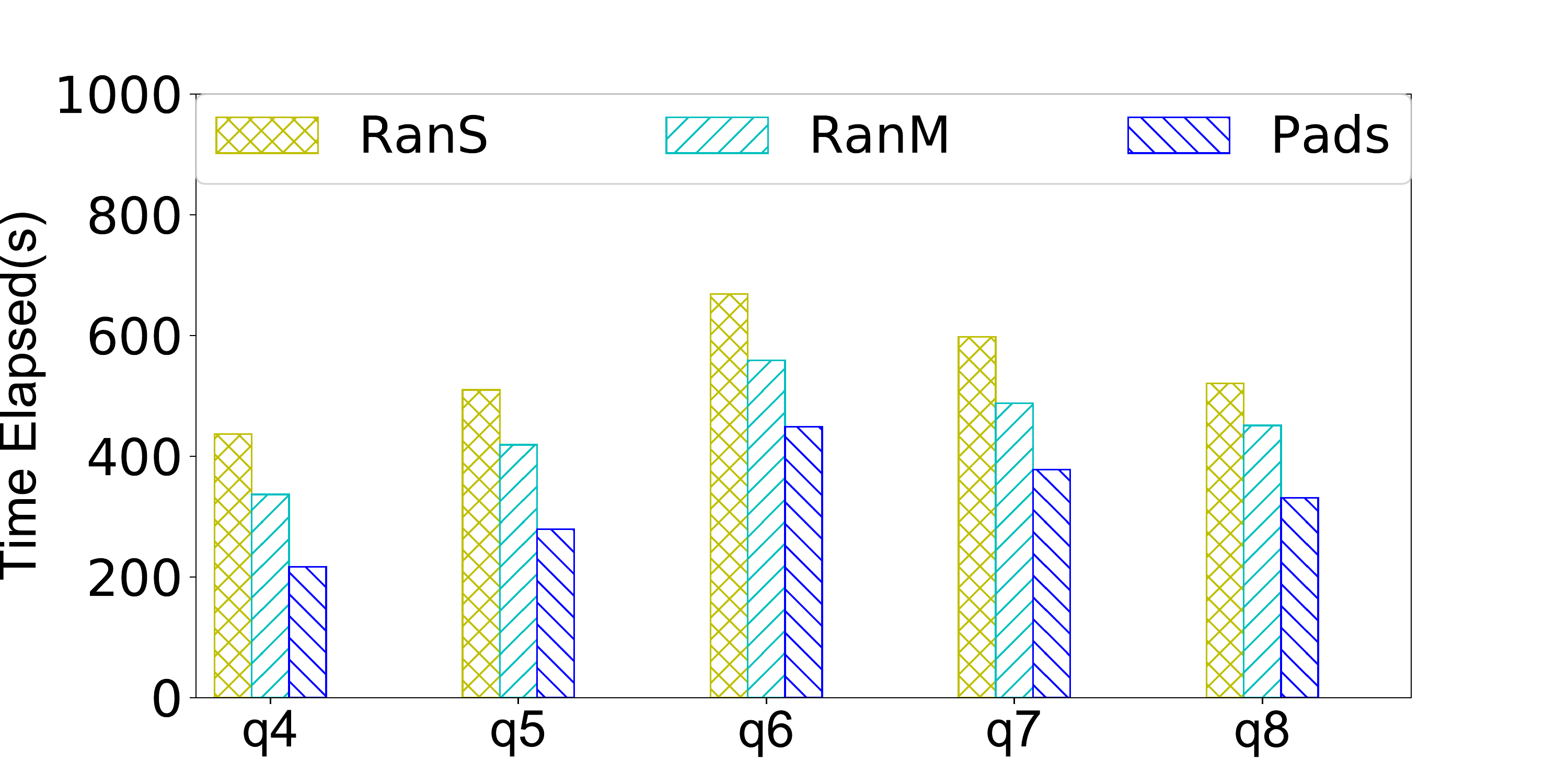}
         \caption{DBLP}
      \end{subfigure}

      \begin{subfigure}[b]{0.25\textwidth}
   		\includegraphics[width=\textwidth]{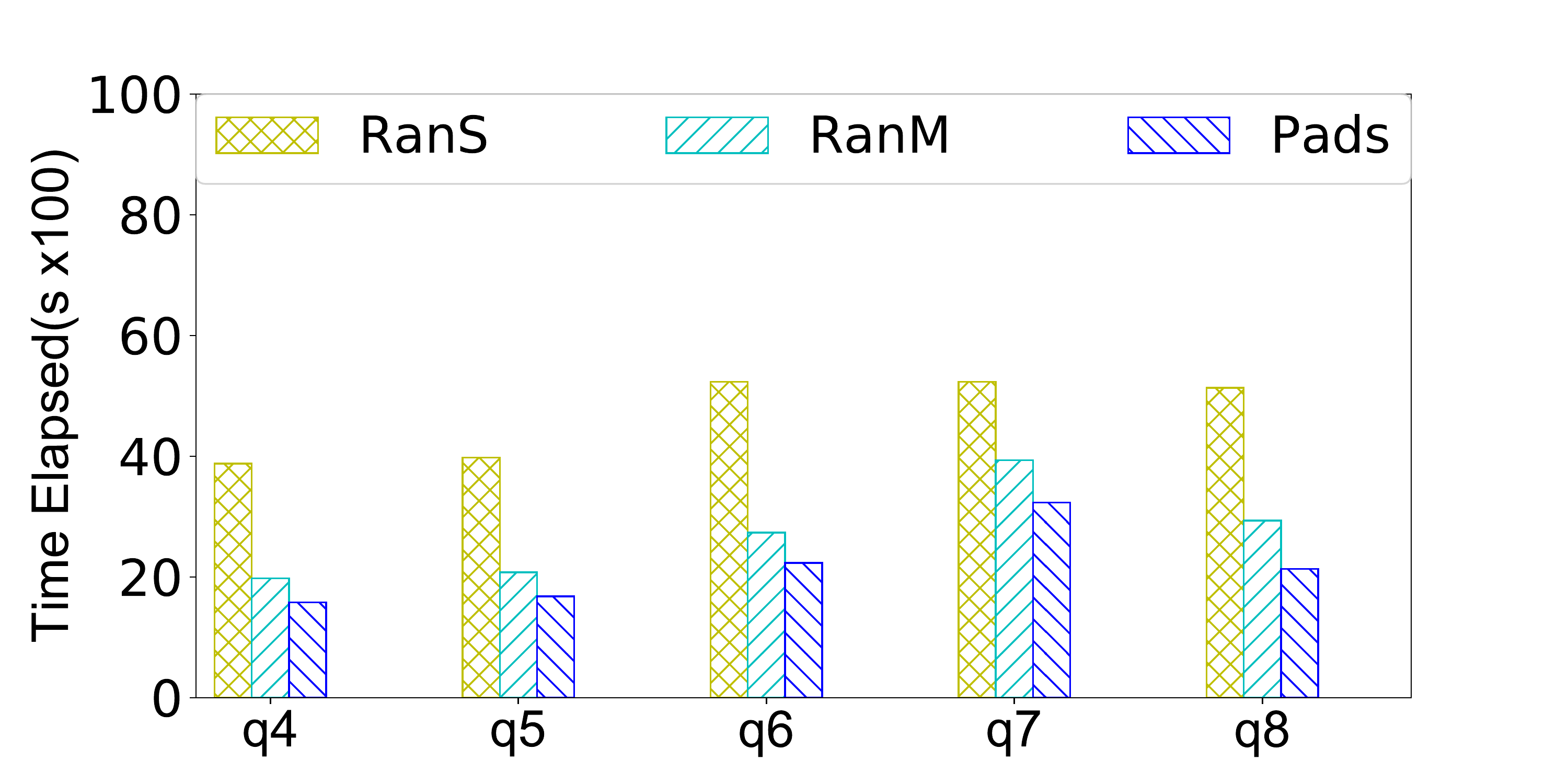}
		\caption{LiveJournal}
      \end{subfigure}%
        ~ 
      \begin{subfigure}[b]{0.25\textwidth}
         \includegraphics[width=\textwidth]{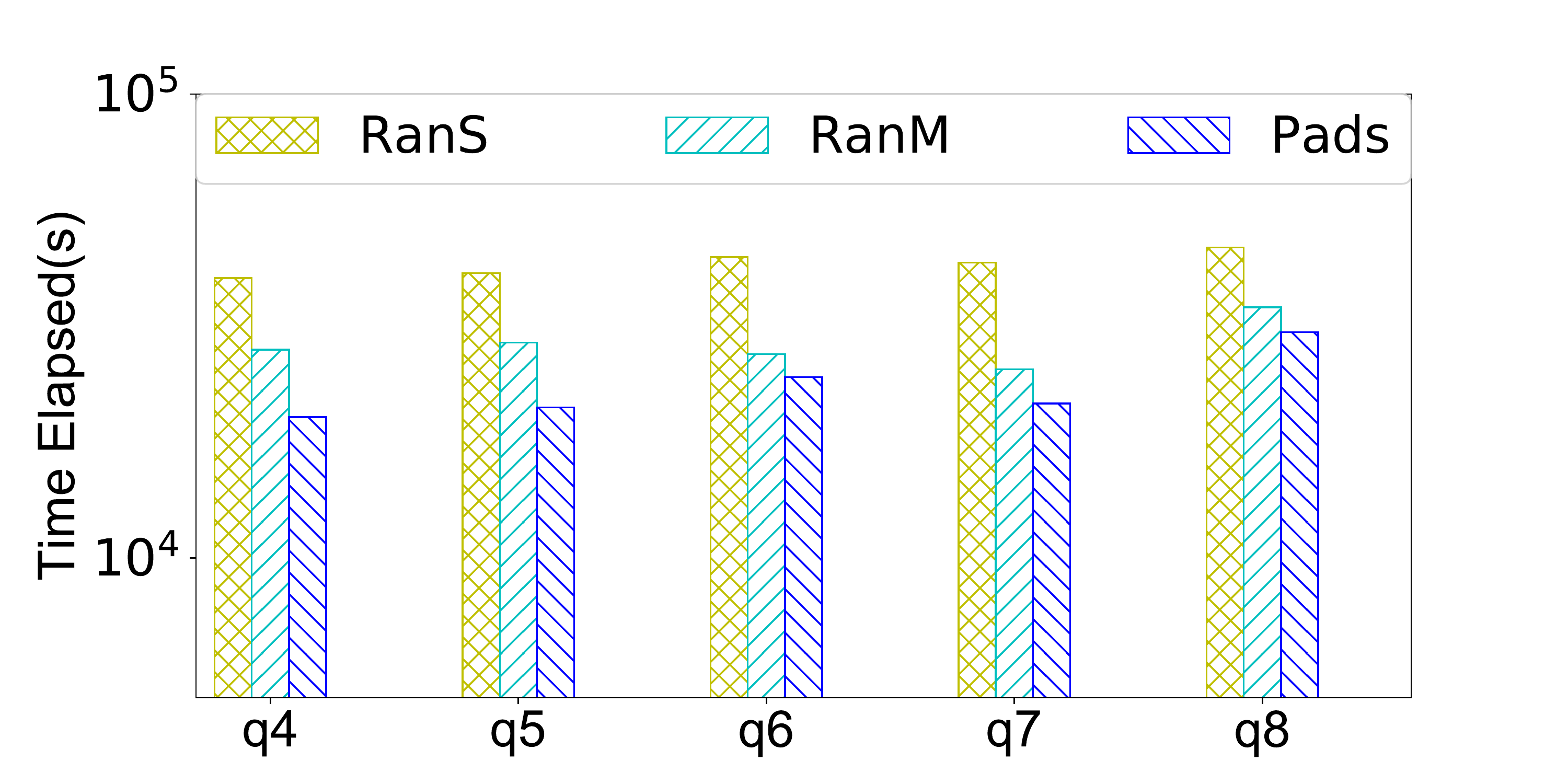}
         \caption{UK2002}
      \end{subfigure}

	\caption{Effectiveness of Execution Plan}
	\label{fig:executionPlanTest}
\end{figure}

The results of Roadnet, DBLP, LiveJournal and UK2002 are as shown in Figure~\ref{fig:executionPlanTest}.  For RoadNet, it is not surprising to see that the processing time are almost the same for the 3 execution plans. This is because most vertices of each RoadNet partition can be processed by \sme, and different distributed query execution plans have little effect over the total processing time. For all other three data sets, it is obvious that our fully optimized execution plan is playing an important role in improving the query processing time, especially when dealing with large graphs such as LiveJournal and UK2002 where large volumes of network communication are generated and can be shared. 

\subsection{Effectiveness of Compression}
To show the effectiveness of our compression strategy, we conducted an experiment to compare the space cost of the simple embedding-list (EL) with that of our embedding trie (ET). We use the RoadNet and DBLP data sets for this test.  The queries are as shown in Figure~\ref{fig:query1}. We omit the test over the other two data sets because the uncompressed volume of the results are too big.


\begin{table}[h]
    \caption{Compression on Roadnet(Mb)}
	\centering
	\small
	\label{t:compression_roadnet}
 	\begin{tabular}{|c|c|c|c|c|c|c|c|c|}
    \hline
     \textbf{Query} & $q_1$ &  $q_2$ &  $q_3$ & $q_4$ & $q_5$ & $q_6$ & $q_7$ &  $q_8$ \\ \hline \hline
   	 \textbf{EL}   & 264   &13 & 65   & 81 & 136   & 183 & -   & -  \\ \hline
     \textbf{ET}	   & 163   & 5 & 33   & 40 & 63   &  73 & -   & -     \\ \hline
	\end{tabular}
\end{table}

\begin{table}[h]
    \caption{Compression on DBLP (Gb)}
	\centering
	\small
	\label{t:compression_dblp}
 	\begin{tabular}{|c|c|c|c|c|c|c|c|c|}
    \hline
     \textbf{Query} & $q_1$ &  $q_2$ &  $q_3$ & $q_4$ & $q_5$ & $q_6$ & $q_7$ &  $q_8$ \\ \hline \hline
   	 \textbf{EL}   & 0.3   & 0.2 & 4.5   & 3.2 & 17.6 & 7.6  & 5.3 & 4    \\ \hline
     \textbf{ET}	   & 0.08M   & 0.06 & 1.1   & 0.7 & 3.8   & 1.3 & 0.9   & 0.8     \\ \hline
	\end{tabular}
\end{table}

The results are as shown in Table~\ref{t:compression_roadnet} and Table~\ref{t:compression_dblp}, respectively. For RoadNet {the intermediate results generated by Queries 7 and 8 are negligible,  therefore they are not listed.}
The results for both datasets demonstrate a good compression ratio.  It is worth noting that the compression ratios of all queries over RoadNet are smaller than that over DBLP. This is because the embeddings of Roadnet are very diverse and they do not share  a lot of common vertices.

\subsection{More Query Processing Results}

\begin{figure}[htbp!]
\small
\centering
\includegraphics[width=0.4\textwidth]{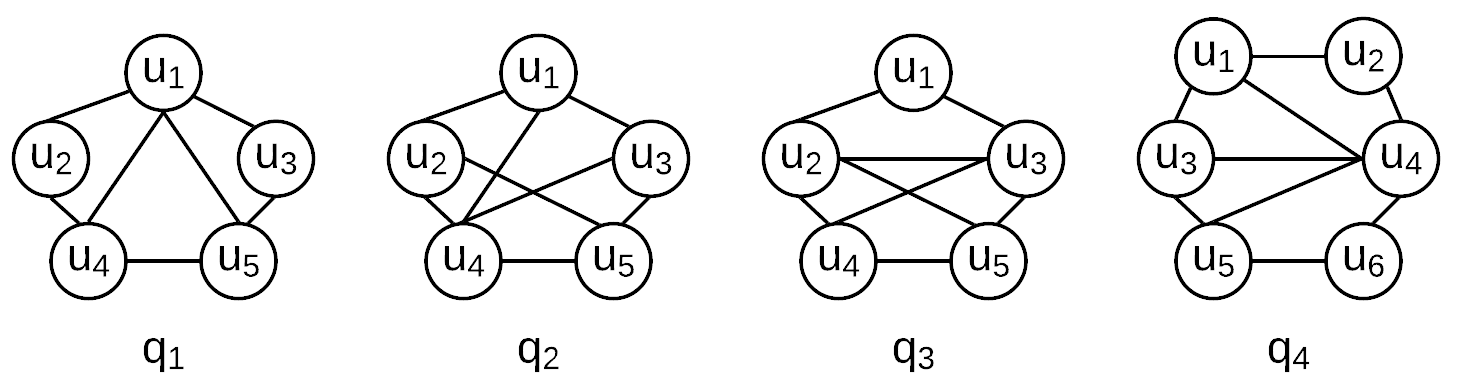}
\caption{Queries with more cliques}
\label{fig:query2}
\end{figure}

As aforementioned, \seed supports clique as decomposition unit and \qiao indexes the cliques in the graph storage. Both methods shall have advantages when processing queries with more cliques. It is noted that most of the queries in Figure~\ref{fig:query1} do not contain any clique. For sound fairness, we also tested some queries from \cite{DBLP:journals/pvldb/QiaoZC17} for the methods of \seed, \qiao and \Pads. The queries are as shown in Figure~\ref{fig:query2}, all of which have cliques. In contrast to the experiment in Section~\ref{subsec:performanceCom}, for \seed, here we also used the program implemented by its original authors. This will guarantee both \seed and \qiao have their maximum optimized performance when processing those queries.

\begin{figure}[htbp!]		
      \centering
      \begin{subfigure}[b]{0.25\textwidth}
   		\includegraphics[width=\textwidth]{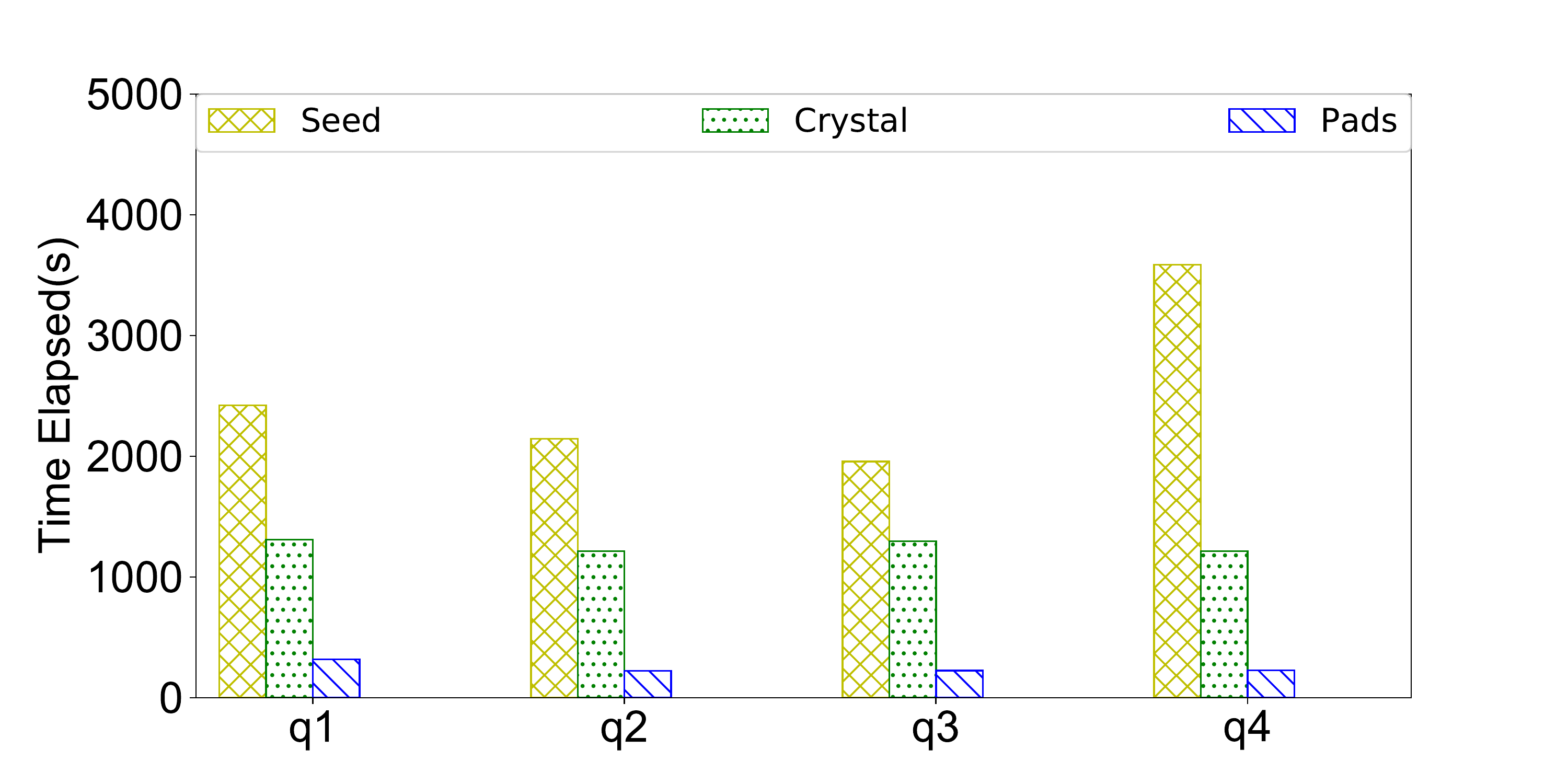}
		\caption{Roadnet}
      \end{subfigure}%
        ~ 
      \begin{subfigure}[b]{0.25\textwidth}
         \includegraphics[width=\textwidth]{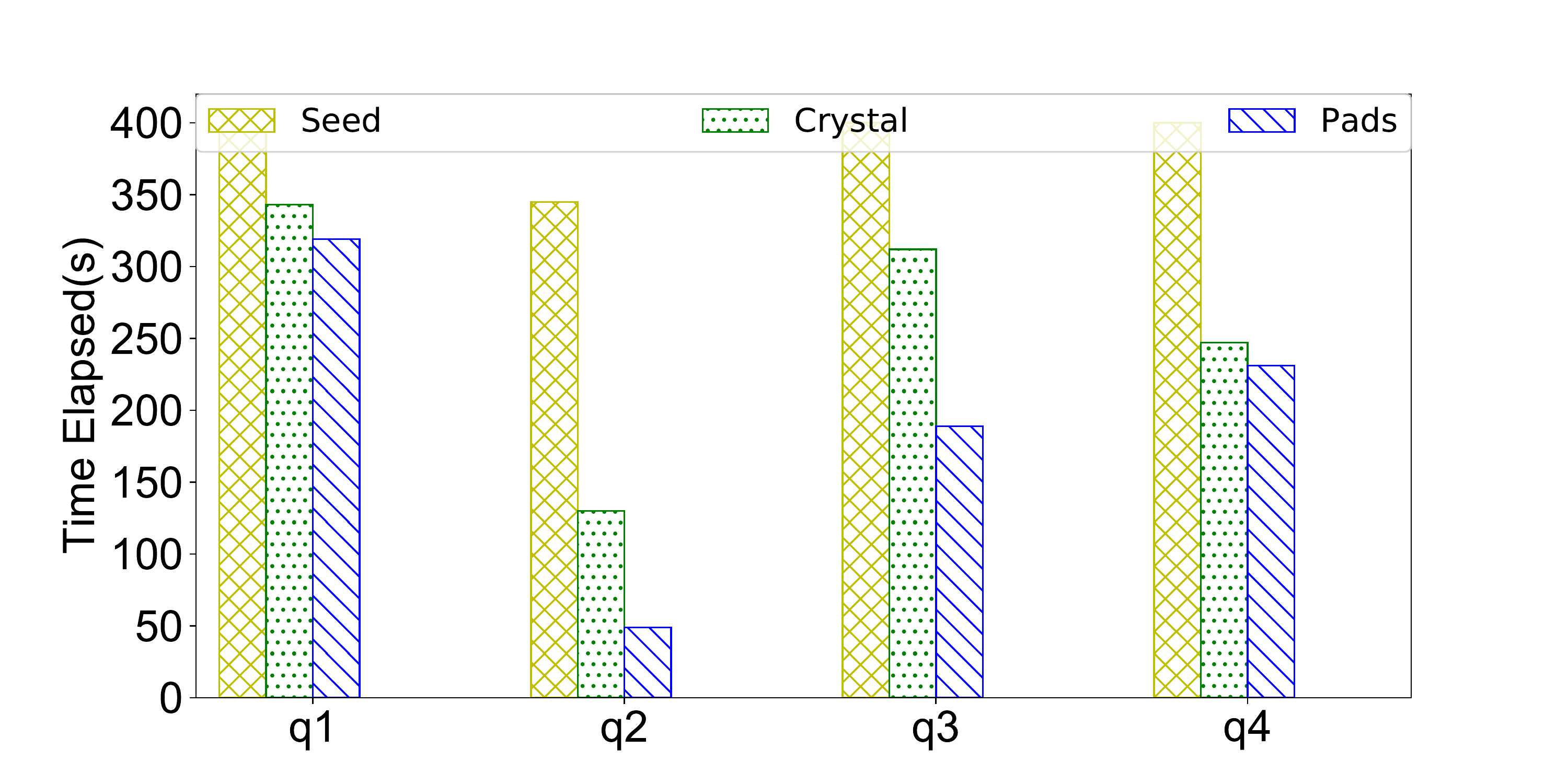}
         \caption{DBLP}
      \end{subfigure}

      \begin{subfigure}[b]{0.25\textwidth}
   		\includegraphics[width=\textwidth]{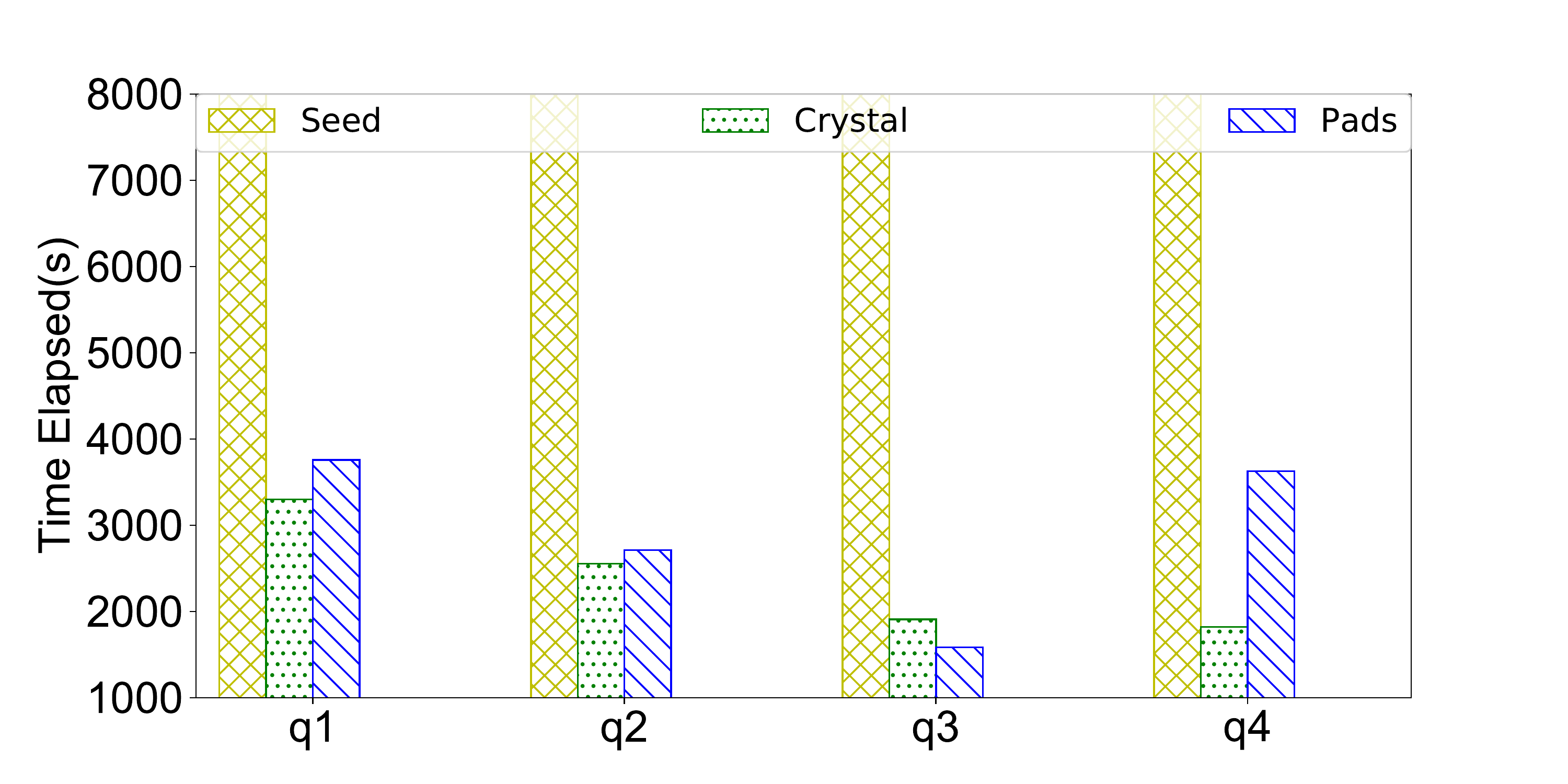}
		\caption{LiveJournal}
      \end{subfigure}%
        ~ 
      \begin{subfigure}[b]{0.25\textwidth}
         \includegraphics[width=\textwidth]{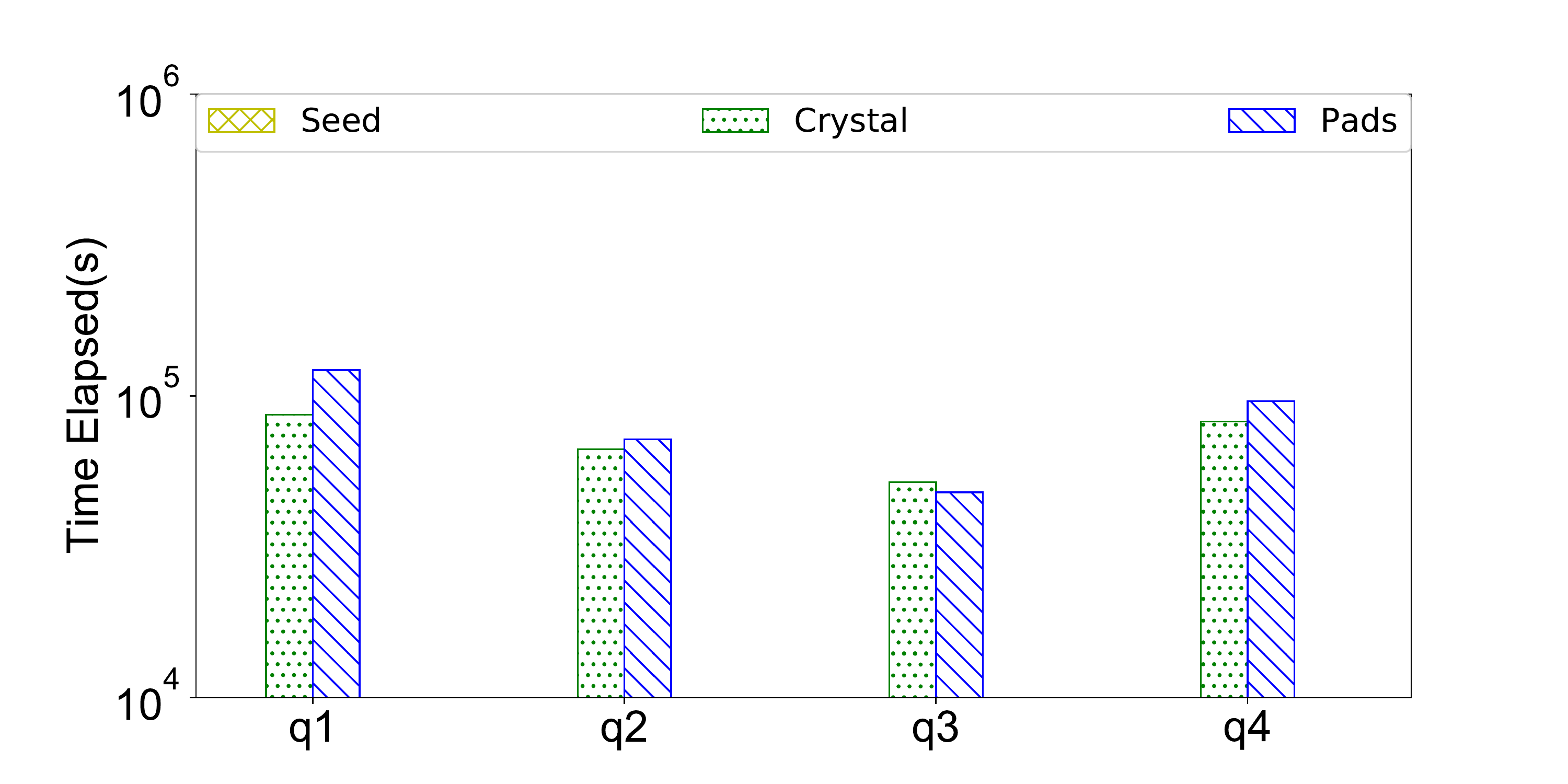}
         \caption{UK2002}
      \end{subfigure}

	\caption{Results of queries with more clique}
	\label{fig:cliqueQuery}
\end{figure}

The results are as shown in Figure~\ref{fig:cliqueQuery}. We omit the results of \seed for UK2002 since its time cost is much higher compared with the other two methods. Being consistent with the result in Section~\ref{subsec:performanceCom}, \Pads performs constantly faster than \seed and \qiao when running on Roadnet (more than 1 order of magnitude) and on DBLP.
For other datasets, \Pads is still better than \seed for all queries, while worse than \qiao for the queries $q_1$, $q_2$ and $q_4$.  This is reasonable because of the heavy clique index of \qiao. However, \Pads has a noticeable improvement over \qiao when processing $q_3$, where the verification edges helped \Pads filtered a lot of unpromising candidates.

\end{document}